%% file: main.tex
\documentclass[12pt, nonatbib]{elsarticle}

\usepackage{amssymb}
\usepackage{amsmath}
\usepackage{caption}
\usepackage[hidelinks]{hyperref}
\usepackage{subcaption}
\usepackage{soul}
\usepackage{color}
\usepackage{setspace}
\renewcommand\hl[1]{#1}
\usepackage{booktabs}
\makeatletter %this line and the following two are necessary to surpress and error related to \c@author, which seems to be defined differently in package and in biber/biblatex
\let\c@author\relax
\makeatother
\usepackage[backend=biber,style=chem-angew,hyperref=true, articletitle=true, doi=true]{biblatex}
\journal{Journal of Magnetic Resonance}

\begin{document}

\begin{frontmatter}

\title{Sub-sampling of NMR Correlation and Exchange Experiments}

\author[1,2,3]{Julian B. B. Beckmann \corref{cor1}}
\ead{jbbeckmann@mgh.harvard.edu}
\author[1]{Mick D. Mantle}
\author[1]{Andrew J. Sederman}
\author[1]{Lynn F. Gladden}

\affiliation[1]{organization={University of Cambridge, Department of Chemical Engineering and Biotechnology},%Department and Organization
            addressline={Philippa Fawcett Drive}, 
            city={Cambridge},
            postcode={CB3 0AS}, 
            country={United Kingdom}}
\affiliation[2]{organization={Harvard Medical School},%Department and Organization
            addressline={25 Shattuck Street}, 
            city={Boston},
            postcode={MA 02115}, 
            country={United States}}
\affiliation[3]{organization={Martinos Centre for Biomedical Imaging, Massachusetts General Hospital},%Department and Organization
            addressline={149 13th Street}, 
            city={Charlestown},
            postcode={MA 02129}, 
            country={United States}}
            
\cortext[cor1]{Corresponding author}  

\begin{abstract}
Sub-sampling is applied to simulated $T_1$-$D$ NMR signals and its influence on inversion performance is evaluated. For this different levels of sub-sampling were employed ranging from the fully sampled signal down to only less than two percent of the original data points. This was combined with multiple sample schemes including fully random sampling, truncation and a combination of both. To compare the performance of different inversion algorithms, the so-generated sub-sampled signals were inverted using Tikhonov regularization, modified total generalized variation (MTGV) regularization, deep learning and a combination of deep learning and Tikhonov regularization. Further, the influence of the chosen cost function on the relative inversion performance was investigated. Overall, it could be shown that for a vast majority of instances, deep learning clearly outperforms regularization based inversion methods, if the signal is fully or close to fully sampled. However, in the case of significantly sub-sampled signals regularization yields better inversion performance than its deep learning counterpart with MTGV clearly prevailing over Tikhonov. Additionally, fully random sampling could be identified as the best overall sampling scheme independent of the inversion method.  Finally, it could also be shown that the choice of cost function does vastly influence the relative rankings of the tested inversion algorithms highlighting the importance of choosing the cost function accordingly to experimental intentions.
\end{abstract}


\end{frontmatter}

\input{US/Sub-Chapters/Undersampling}

\printbibliography[title=References]

\end{document}

%% file: US/Sub-Chapters/Undersampling.tex
\section{Introduction}
\label{sec:US_int}

Research interest into signal processing dates back to the early days of Shannon and Nyquist and the then developed Nyquist-Shannon theorem\supercite{Shannon1949CommunicationNoise, Shannon1948ACommunication} provided the theoretical basis for the nowadays omnipresent signal digitization. Due to the high measurement time and memory requirements for experiments that fulfill the Nyquist-Shannon theorem, research focus shifted quickly towards methods allowing for lower sampling rates as Nyquist-Shannon, but preventing artifacts associated with non-fully sampled signals such as aliasing.\supercite{Brunton2019Data-DrivenEngineering} The most major breakthrough in recent years towards this field of research was the development of compressed sensing.\supercite{Candes2006StableMeasurements, Candes2006RobustInformation, Massa2015CompressiveReview, Donoho2009Message-passingSensing} Compressed sensing makes use of the sparsity of many signals in some domain.\supercite{Brunton2019Data-DrivenEngineering, Candes2006RobustInformation, Candes2006StableMeasurements} For instance, magnetic resonance images are sparse as well as centered in k-space and hence, the k-space representation can be truncated allowing for a compressed representation.\supercite{Holland2014LessChemistry} In the field of magnetic resonance, this property can be employed to significantly reduce the measurement time during MRI or NMR spectroscopy experiments.\supercite{Lustig2008CompressedMRI, Holland2011FastSensing, Zhang2014EnergyMRI} This is usually achieved by using a sampling pattern which focuses on the \hl{centre} of k-space, allowing for a compression of more than $90\%$ in ideal cases compared to the fully sampled signal.\supercite{Lustig2008CompressedMRI, Benning2014PhaseApproaches, Holland2014LessChemistry} The concept of compressed sensing can then be applied to reconstruct the original image.\supercite{Candes2006StableMeasurements, Candes2006RobustInformation, Holland2014LessChemistry} This procedure is well established for a vast number of MRI experiments but is less commonly applied to other type of NMR experiments. Hence, the focus of this study is to apply the idea of signal compression to NMR correlation and exchange experiments and to provide an analysis regarding the possible level of sub-sampling without significant loss of detail or the introduction of artifacts. It is further analysed how truncation compared to random sampling \hl{affects} the performance of the reconstruction procedure.

\section{Theoretical Background}
\label{sec:US_theo}

In the case of magnetic resonance imaging, the reconstruction of the original image from a sub-sampled k-space is described by the following least-squares optimization problem:\supercite{Candes2006StableMeasurements, Candes2006RobustInformation, Holland2014LessChemistry, Lustig2008CompressedMRI, Benning2014PhaseApproaches}
\begin{equation} \label{eq:US_cs}
    \underline{\mathbf{F}} = \arg \; \min {}_{\underline{\mathbf{F}} \, \geq \, 0} \; \left|\left| \underline{\mathbf{R}} \, \underline{\mathcal{F}}^{-1} \, \underline{\mathbf{F}} - \underline{\mathbf{R}} \, \underline{\mathbf{S}} \right| \right|_{2}^{2}
\end{equation}
where for an imaging experiments $\underline{\mathbf{F}}$ coincides with the image, $\underline{\mathbf{S}}$ represents a fully sampled k-space, $\underline{\mathcal{F}}$ is the Fourier matrix and $\underline{\mathbf{R}}$ refers to the sub-sampling matrix defining the sampling pattern employed for k-space acquisition. Using $\tilde{\underline{\mathbf{K}}} = \underline{\mathbf{R}} \, \underline{\mathcal{F}}^{-1}$ and $\tilde{\underline{\mathbf{S}}} = \underline{\mathbf{R}} \, \underline{\mathbf{S}}$, equation~\ref{eq:US_cs} transforms to:
\begin{equation} \label{eq:US_cs_trans}
    \underline{\mathbf{F}} = \arg \; \min {}_{\underline{\mathbf{F}} \, \geq \, 0} \; \left|\left| \tilde{\underline{\mathbf{K}}} \, \underline{\mathbf{F}} - \tilde{\underline{\mathbf{S}}} \right| \right|_{2}^{2}
\end{equation}
This is an identical optimization problem compared to the inversion of the signal stemming from a NMR correlation or exchange experiment. In this case, $\tilde{\underline{\mathbf{K}}}$ represents the kernel matrix which is usually given by some sort of multivariate exponential decay with its exact definition depending on the employed pulse sequence, $\tilde{\underline{\mathbf{S}}}$ coincides with the time or gradient encoded NMR signal and $\underline{\mathbf{F}}$ with the to-be-estimated distribution such as a $T1$-$D$ correlation map. Hence, the concept of compressed sensing can be applied to NMR correlation and exchange experiments without the necessity of adapting the employed inversion equation. Here, compressed sensing can just be understood to sub-sample the NMR signal compared to a threshold signal which is considered to be fully sampled. In this study, the fully sampled signal is considered to be the signal with an equal number of points as the reconstructed distribution, whereas every signal with less points as the reconstructed distribution is referred to as sub-sampled. 

\section{Methodology}
\label{sec:US_meth}

The objective of this paper is to evaluate the reconstruction performance of sub-sampled signals from NMR correlation and exchange experiments with their fully sampled counterparts. From an experimental point of view, the optimal sub-sample scheme can be considered to result in a significant reduction of acquisition time while keeping the quality of inversion close to constant compared to a fully sampled reconstruction. Hence, a comparison of reconstruction quality depending on the method of inversion and the sub-sample scheme is expected to establish a \hl{rationale} for choosing the best combination of inversion method as well as sampling scheme. Consequently, in this paper the quality of reconstruction using different inversion algorithms as well as different sub-sampling methods are compared. In particular, for signal inversion Tikhonov regularization,\supercite{Tikhonov1978SolutionsProblems., Brunton2019Data-DrivenEngineering, Golub1999TikhonovSquares} MTGV regularization,\supercite{Reci2017RetainingExperiments., Beckmann_MTGV} deep learning\supercite{Beckmann_dl} and a combination of Tikhonov regularization and deep learning\supercite{Beckmann_dl} are employed, whereas sub-sampling is done randomly, via truncation and via a combination of truncation and random sampling. 
\begin{figure}[t]
    \centering
    \begin{subfigure}[b]{0.475\textwidth}
        \centering
        \includegraphics[width=\textwidth, keepaspectratio]{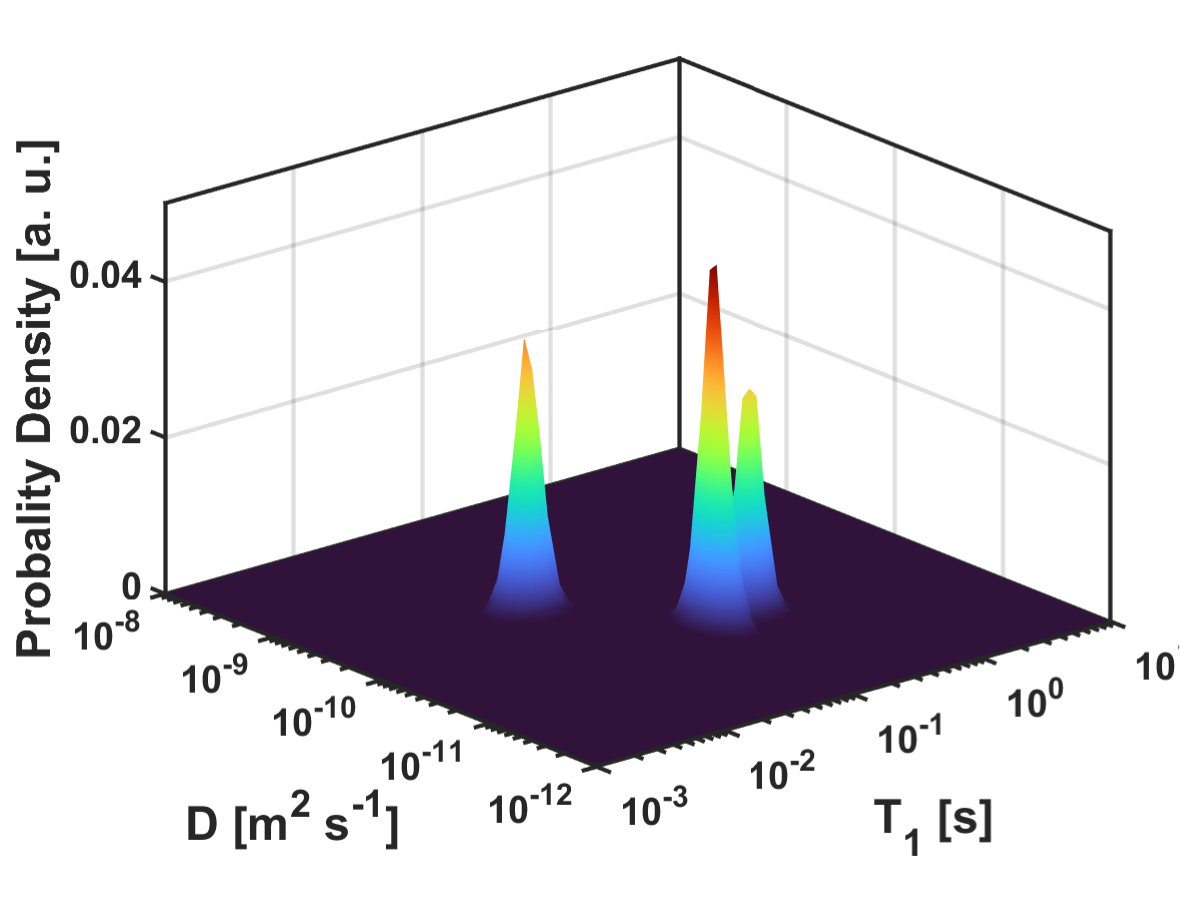}
        \caption{Sparse components only}
        \label{fig:dis_real_sparse}
    \end{subfigure}
    \begin{subfigure}[b]{0.475\textwidth}
        \centering
        \includegraphics[width=\textwidth, keepaspectratio]{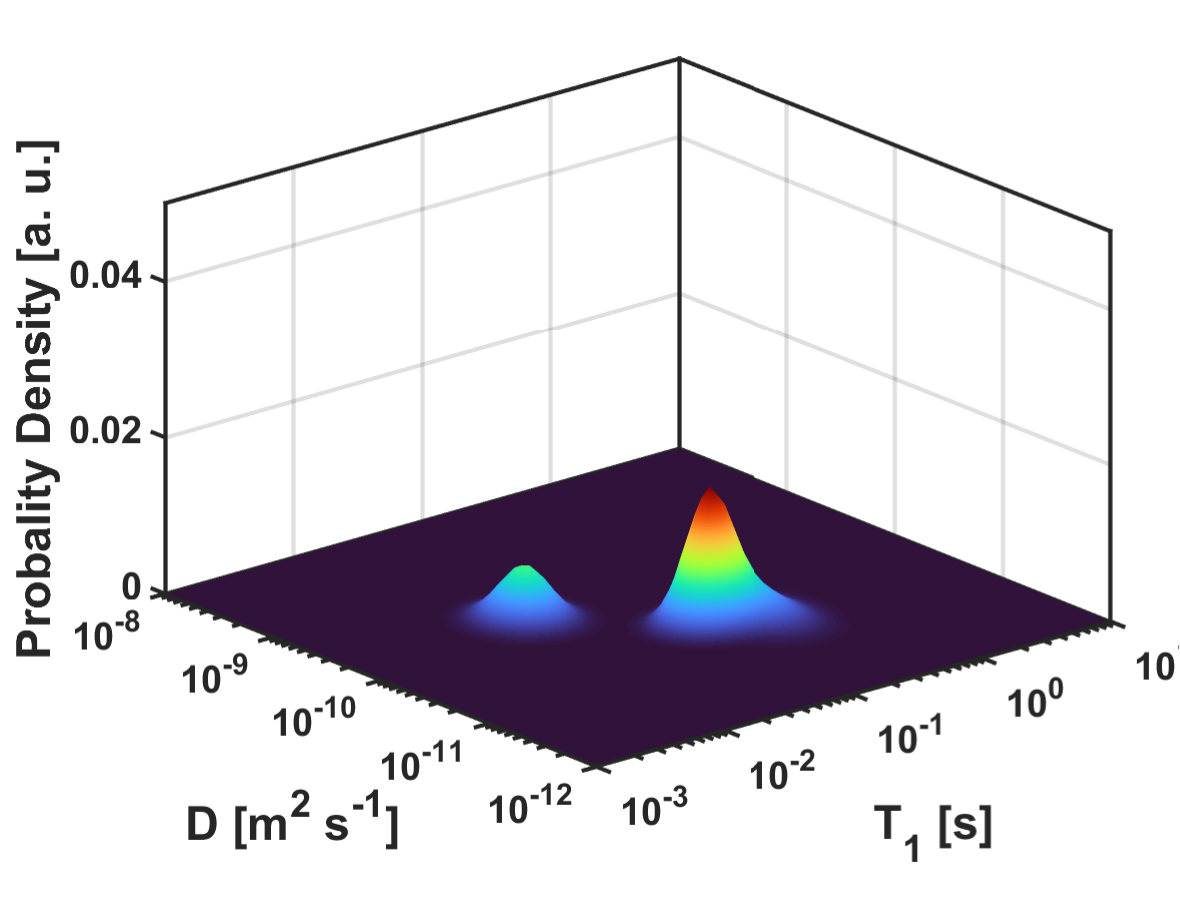}
        \caption{Smooth components only}
        \label{fig:dis_real_smooth}
    \end{subfigure}
    \\
    \begin{subfigure}[b]{0.475\textwidth}
        \centering
        \includegraphics[width=\textwidth, keepaspectratio]{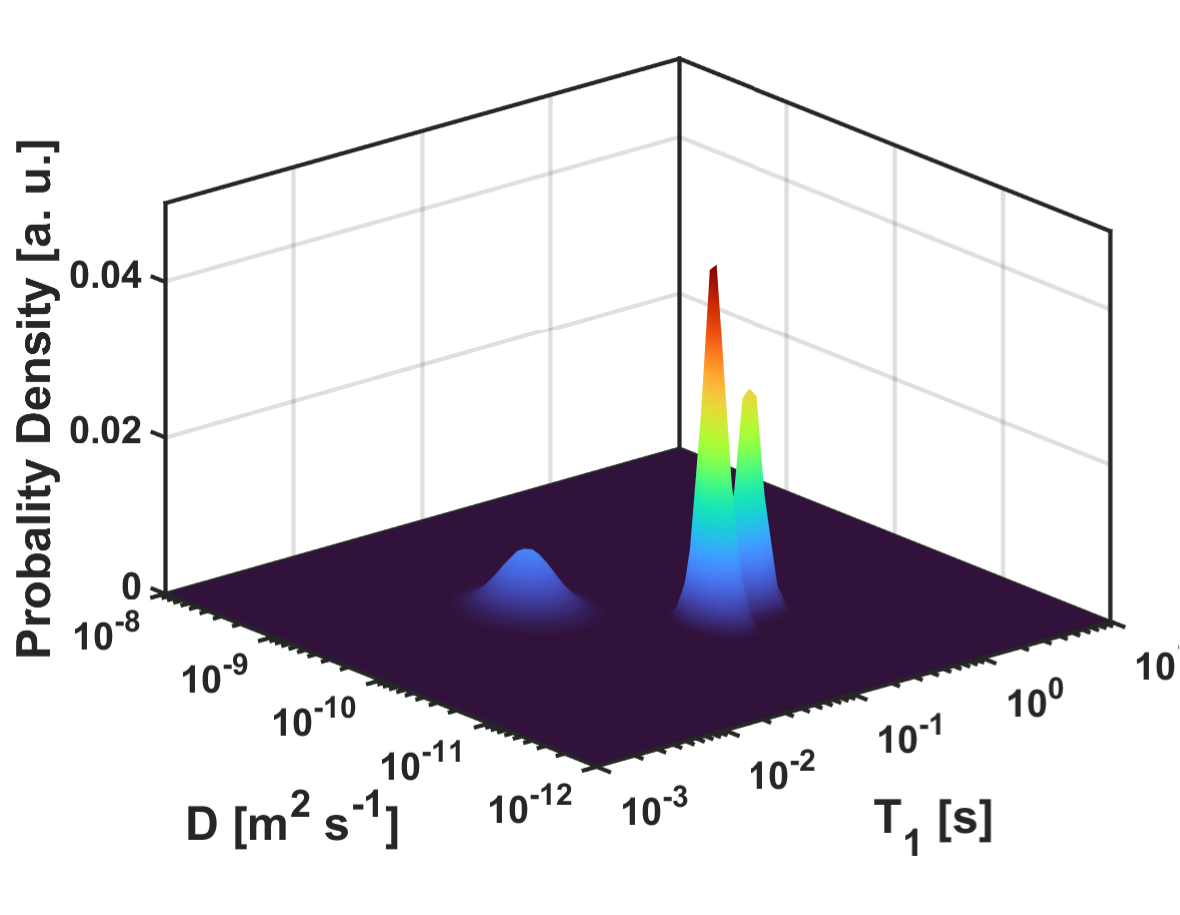}
        \caption{One smooth and two sparse components}
        \label{fig:dis_real_sm_sp}
    \end{subfigure}
    \caption{$T_1$-$D$ distributions employed in this publication.}
    \label{fig:US_dis_sim}
\end{figure}
To evaluate the reconstruction performance of all possible sub-sampling and inversion permutations, three different $T_1$-$D$ distributions are simulated. In this case, $T_1$-$D$ experiments are chosen as the objective of study because compared to other NMR correlation or exchange experiments such as $D$-$T_2$ or $T_2$-$T_2$, $T_1$-$D$ measurements suffer from particularly slow acquisition times and hence, $T_1$-$D$ experiments benefit over-proportionally from sub-sampling. However, findings made for $T_1$-$D$ are expected to be equally applicable to other type of NMR correlation and exchange experiments due to the translational invariance of the inversion methods employed. All employed distributions span 64 logarithmically spaced data points in both directions, contain three components and are given as a superposition of two-dimensional log-normal distributions. Peak positions are identical for all distributions considered, but different levels of sparsity and smoothness are given, with two out of three distributions containing solely smooth or sparse components and the final distribution consisting of a mixture between sparse and smooth peaks. 
\begin{table}[t]
\centering
\caption{Lognormal parameters of the employed $T_1\mbox{-}D$ distributions. The labels~a, b and c coincide with the labelling in figure~\ref{fig:US_dis_sim}.}
\label{tab:dis_par_us}
\resizebox{\textwidth}{!}{
\begin{tabular}{ccccc}
\toprule
Distribution & $T_1$ log mean      & $T_1$ log standard deviation & $D$ log mean          & $D$ log standard deviation \\ \midrule
a            & -0.84; -0.56; -1.52 & 0.05; 0.08; 0.09             & -10.51; -10.51; -9.60 & 0.10; 0.06; 0.07            \\
b            & -0.84; -0.56; -1.52 & 0.13; 0.15; 0.15             & -10.51; -10.51; -9.60 & 0.15; 0.26; 0.18           \\
c            & -0.84; -0.56; -1.52 & 0.05; 0.08; 0.15             & -10.51; -10.51; -9.60 & 0.10; 0.06; 0.18            \\ \bottomrule
\end{tabular}%
}
\end{table}
The minimum diffusion coefficient considered in the distribution is $10^{-12}\, \mathrm{\frac{m^2}{s}}$ and the maximum coincides with $10^{-8}\, \mathrm{\frac{m^2}{s}}$, whereas for $T_1$, the distribution boundaries are $10^{-3}\, \mathrm{s}$ and  $10^{1}\, \mathrm{s}$ respectively. The resulting distributions are given in figure~\ref{fig:US_dis_sim} and their log-normal parameters can be found in table~\ref{tab:dis_par_us}. From this, the simulated signal is obtained by multiplying the distribution with the $T_1$-$D$ kernel matrix whose elements are given by the following exponential decay:
\begin{equation}
    k(t, g) = \left(1 - \exp{\left(- \frac{t}{T_1}\right)}\right) \exp{\left(- q\left(g\right) D \right)},
\end{equation}
where $t$ refers to the time delays employed for $T_1$ encoding and $q\left(g\right)$ is a quadratic function depending on the gradient $g$ which exact definition is subject to the pulse sequence used for diffusion encoding.\supercite{PaulCallaghan2011TranslationalResonance, Tanner1970UseStudies, Cotts1989PulsedSystems, Stejskal1965SpinGradient} To the received signal, Gaussian noise is added to match signal to noise ratios of 20, 200 and 2000. The final signal spans 64 $\times$ 64 logarithmically spaced data points with the $T_1$-encoding time delays constraint to the identical boundaries as it is the case for the $T_1$ dimension of the $T_1$-$D$ distributions and the diffusion encoding boundaries given by the inverse of the diffusion coefficient upper and lower limit. From those fully sampled signals, sub-sampling with an increment of eight data points is employed. Hence, subsets spanning 56 data points in both directions down to $8 \times 8$ data points are generated and the resulting signals are used as an initial input for the reconstruction methods considered. In the case of Tikhonov regularization, the algorithm described by Mitchell \textit{et al.}\supercite{Mitchell2012NumericalDimensions} is employed, whereas MTGV regularization\supercite{Reci2017RetainingExperiments.} in combination with generalized cross-validation\supercite{Golub1979GeneralizedParameter, Xu2017PredictedPrediction} as described in an earlier publication of Beckmann and co-workers is used.\supercite{Beckmann_MTGV} For reconstructions via deep learning directly from the input signal, the procedure described in Beckmann \textit{et al.}\supercite{Beckmann_dl} was followed with the amendment that sub-sampled signals were interpolated to 64 $\times$ 64 data points using modified Akima piece-wise cubic Hermite interpolation.\supercite{Akima1970AProcedures, Akima1974AProcedures} This is necessary due to the fixed input size of the employed neural network. The interpolation boundaries are chosen to coincide with the lower and upper limits of the sub-sampled signals and the interpolation method is then used to estimate 64 $\times$ 64 logarithmically spaced data points fulfilling the latter boundary conditions. Eventually, a reconstructed distribution is obtained by passing the interpolated signal to the neural network. In the case of the combined approach between Tikhonov regularization and deep learning, the reconstruction results obtained from Tikhonov regularization are passed to a neural network which acts as a filter intending to remove reconstruction artefacts. The neural network structure employed here is identical to the method developed in previous work of Beckmann and co-workers\supercite{Beckmann_dl} with the difference that the network is not trained on pairs of artificial signals and their associated distributions instead the simulated signals are firstly processed via Tikhonov regularization and the obtained reconstructions are used in combination with the real distributions as input and output for network training. \hl{Hence, network training follows an analogous procedure as described in Beckmann \textit{et al.}{\supercite{Beckmann_dl}} with the exception that the reconstructions obtained from Tikhonov regularization are employed as inputs instead of the simulated NMR signals. Consequently,} the reader is referred to the methods section of the deep learning paper of Beckmann and co-workers for exact technical details on network structure, training and method employment.\supercite{Beckmann_dl} The benefit of this combined approach is twofold. Firstly, pre-processing of the sub-sampled signals via Tikhonov regularization is used to match the size of the initial reconstruction with the input size of the neural network rendering the interpolation step prior to the neural network unnecessary. Secondly, the neural network acts as a filter removing artifacts such as non-existing peaks from the initial reconstruction. This procedure could also be employed with MTGV instead of Tikhonov regularization. However, the processing time for MTGV regularization is commonly around one order of magnitude higher \hl{than is} the case for Tikhonov regularization.\supercite{Reci2017RetainingExperiments.} This is usually acceptable if only a small number of reconstructions has to be processed, but for network training more than $10^5$ training samples are normally necessary to reach sufficient network performance, hence rendering the simulation of sufficient training data unfeasible. Consequently, this paper solely focuses on the combination of Tikhonov pre-processing and final reconstruction via deep learning. Those reconstruction methods are then combined with previously mentioned sub-sampling methods. \hl{In the case} of random sub-sampling, a subset is generated by randomly sampling data points from the fully sampled signal until the maximum number of points in the subset is reached. For the purpose of truncation, the first $N_1$ $\times$ $N_2$ data points (with the first data point being associated with maximum signal intensity) are kept and the rest discarded, where $N_1$ and $N_2$ are the dimensions of the subset. If instead a combination of truncation and random sampling is employed, the first $4 \times 4$ data points are kept and the remaining data points in the subset are obtained via random sampling. The last sub-sampling method is inspired by the usage of compressed sensing in the field of magnetic resonance imaging. \hl{During imaging} experiments it is common to fully sample the \hl{centre} of k-space, but to employ only sparse random sampling in the remaining part of k-space.\supercite{Holland2014LessChemistry} Hence, the full sampling of the first 4 $\times$ 4 data points is intended to acquire the data points with the highest signal intensity which is then combined with random sampling for the remaining data points in the subsets analogously to an imaging experiment. To evaluate the performance of the inversion methods and the sub-sampling procedure, a ranking method has to be introduced. Hence, a modified version of the cost function used in the MTGV work of Reci \textit{et al.}\supercite{Reci2017RetainingExperiments.} is employed to compare different \hl{reconstruction} results with a score of zero coinciding with a perfect reconstruction. The resulting equation is given by the following expression:
\begin{equation} \label{eq:score_us}
    \chi = \sum_i \frac{\left(f_{i, true} - f_{i, rec} \right)^2}{\max \left(10^{-4}, \min \left(f_{i,rec}, f_{i, true} \right) \right)},
\end{equation}
where $f_{i, true}$ and $f_{i, rec}$ coincide with the $i$-th element of the real and reconstructed distribution respectively. This equation only differs compared to the cost function employed by Reci \textit{et al.}\supercite{Reci2017RetainingExperiments.} by the introduction of an additional $\min$-term in the denominator. The idea behind this modification is to increase the penalty for reconstructions which are significantly too sparse compared to the real distributions. For instance, a broad smooth peak could be split up into multiple components, which would not be heavily penalized through the cost function originally used by Reci and co-workers.\supercite{Reci2017RetainingExperiments.} If instead equation~\ref{eq:score_us} is employed, points in the reconstruction close to zero in between two falsely split up peaks is assigned a higher penalty, because $\min \left(f_{i,rec}, f_{i, true} \right)$ in the denominator ensures that the least squares term is divided by the smallest possible value, hence increasing the penalty. Overall, equation~\ref{eq:score_us} allows for a compromise between fidelity and the suppression of reconstruction artefacts. In more detail, reconstruction artefacts such as additional non-existing peaks or the partition of smooth components into multiple peaks is penalized more heavily but the magnitude of the additional penalty is chosen small enough to prevent the suppression of the fidelity effect originating from the least squares contribution in equation~\ref{eq:score_us}. However, from \hl{an} experimental point of view, it is usually not the case that the real distribution is known and \hl{hence} artefact peaks can hinder the clear interpretation of an experiment significantly. In addition, for many applications identifying the correct number of components as well as accurate estimates for the logarithmic means of the distributions is more important than good approximations of the variance and peak shape. Hence, a ranking method which penalizes artefact peaks very heavily can be beneficial for the clear interpretation of experimental results even if the high artefact penalty dominates penalty contributions stemming from considerable variations in variance and peak shape. A penalty function which provides strong emphasis on the suppression of artefacts can be defined as follows:
\begin{equation} \label{eq:score_us_art}
    \phi = \sum_i w_i \left(f_{i, true} - f_{i, rec} \right)^2 \; \mathrm{with} \; w_i = 
    \begin{cases}
    \frac{10^4}{\max \left( \underline{\mathbf{F}} \right)},  f_{i, true} < 10^{-4} \\
    1,  f_{i, true} \geq 10^{-4}
    \end{cases}, 
\end{equation}
with $\underline{\mathbf{F}}$ being the reconstructed distribution. This function provides a very significant penalty whenever the real distribution is close to zero and at the same time the reconstructed distribution deviates considerably from zero. Consequently, reconstructions containing artefact peaks are penalized very severely which is beneficial for the unambiguous interpretation of experimental results. Moreover, the weights are scaled based on the maximum peak height of the reconstruction considered, which provides a relative penalty rather than an absolute one. This is beneficial because, \hl{in the case of a very sparse reconstruction with high intensity peaks,} a small artefact peak results in less ambiguity than a peak of the same height \hl{would generate} in a smooth reconstruction with overall lower peak intensities. Hence, the penalty in the sparse case should be lower compared to an artefact peak of identical height in a smooth reconstruction, which is exactly achieved by the relative weighting employed in equation~\ref{eq:score_us_art}. Eventually, ranking all reconstructions with both prior-discussed penalty functions does not only allow for the identification of the best combination of sub-sampling procedure and reconstruction method, but is also expected to provide additional insight regarding the influence of the chosen penalty function on the overall rankings of sub-sampling schemes and inversion methods.

\section{Results and Discussion}
\label{sec:results_us}

\begin{figure}[t!]
    \centering
    \begin{subfigure}[b]{0.45\textwidth}
        \centering
        \includegraphics[width=\textwidth, keepaspectratio]{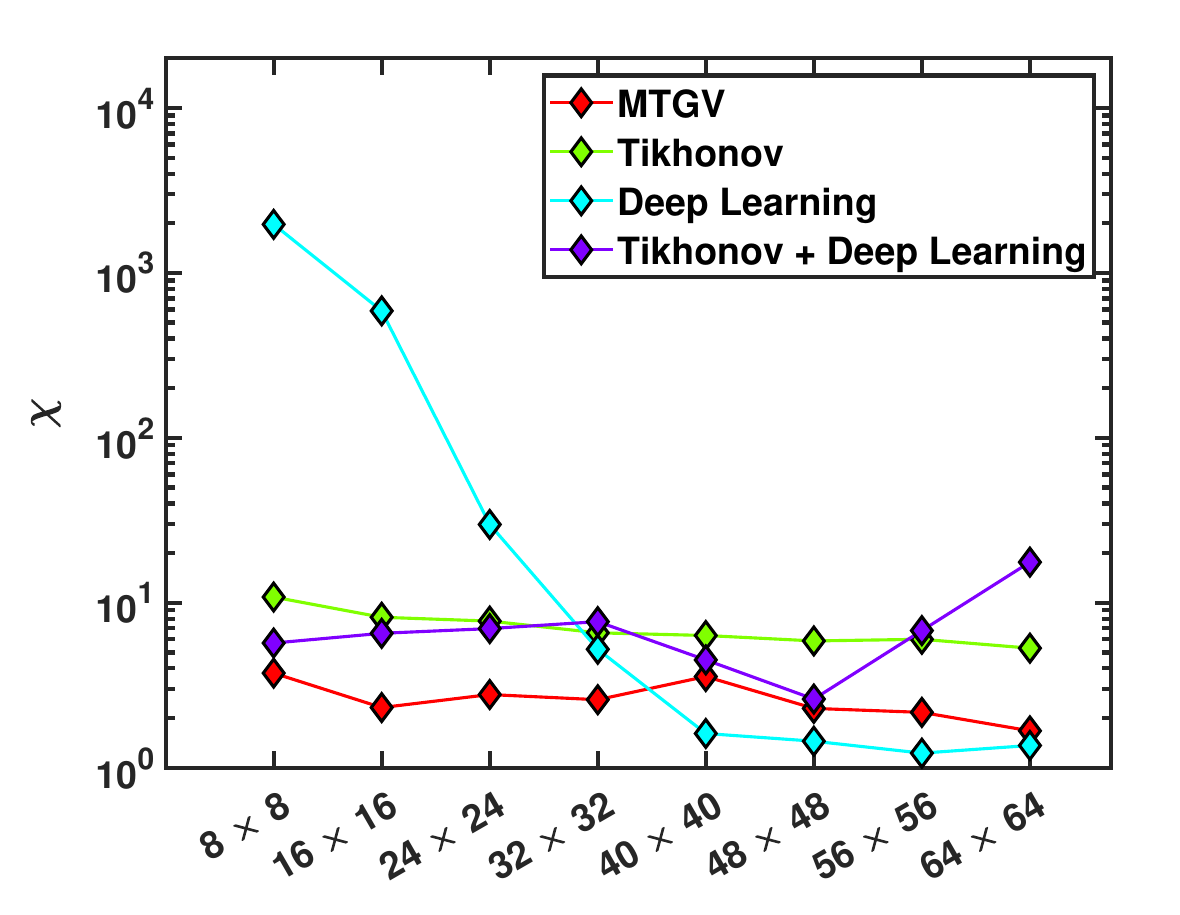}
        \caption{Sparse, random sampling}
        \label{fig:xi_2000_sparse_rs}
    \end{subfigure}
    \hspace{0.01\textwidth}
    \begin{subfigure}[b]{0.45\textwidth}
        \centering
        \includegraphics[width=\textwidth, keepaspectratio]{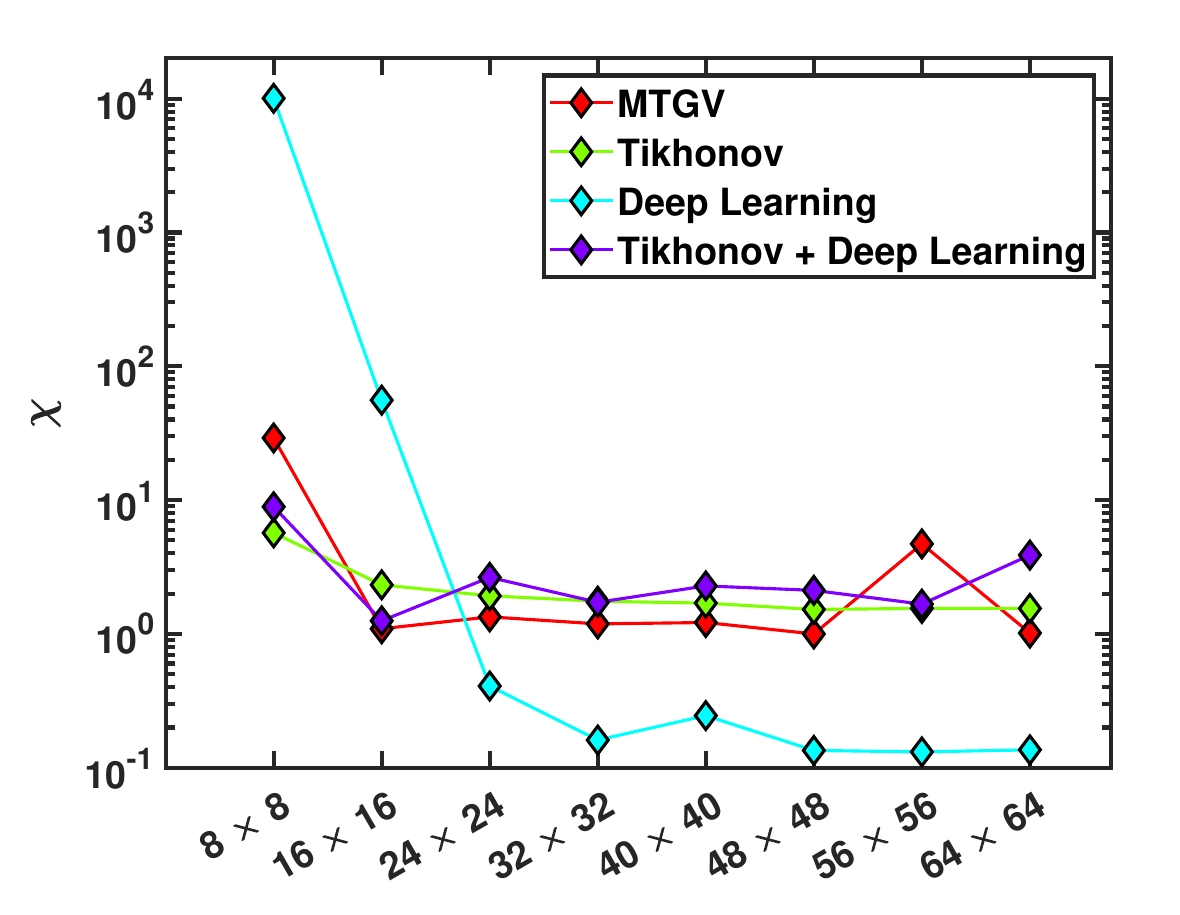}
        \caption{Smooth, random sampling}
        \label{fig:xi_2000_smooth_rs}
    \end{subfigure}
    \\
    \begin{subfigure}[b]{0.45\textwidth}
        \centering
        \includegraphics[width=\textwidth, keepaspectratio]{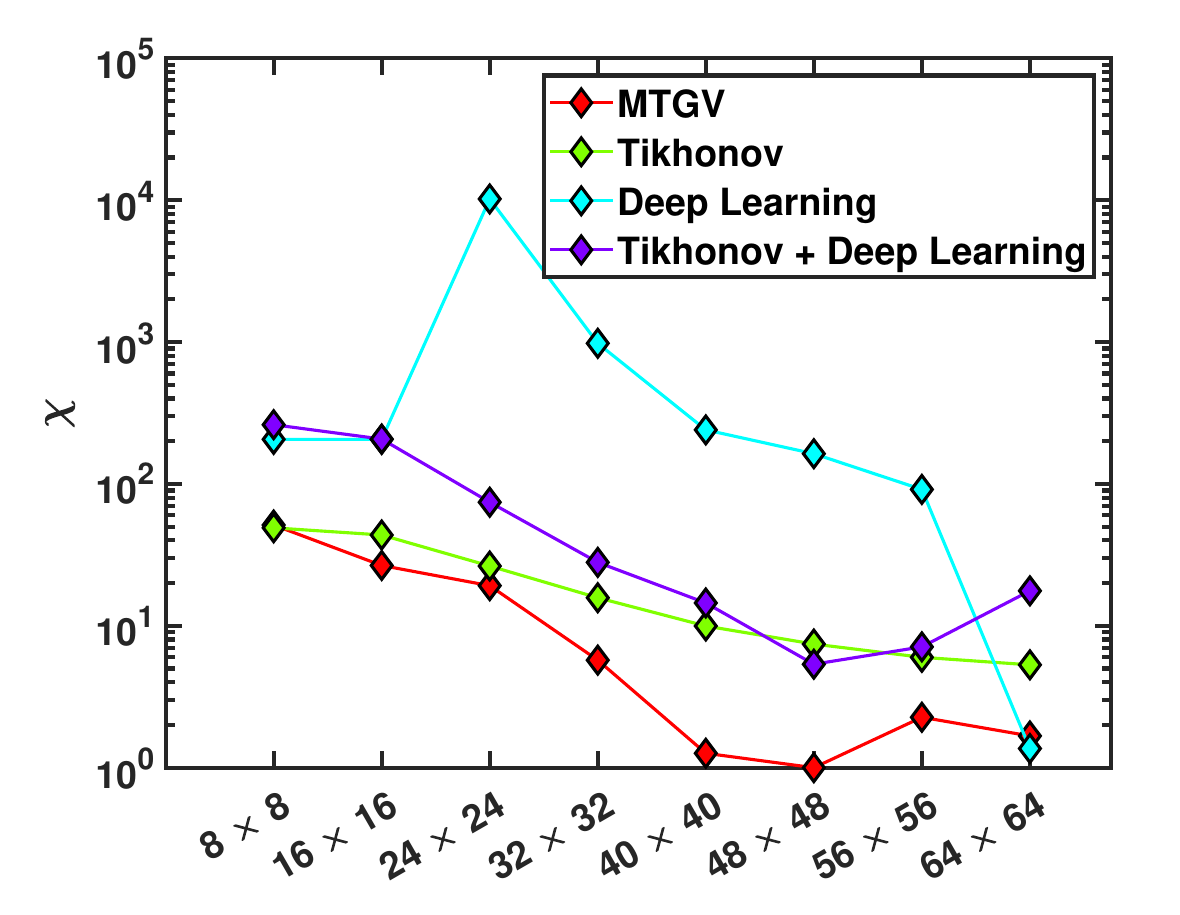}
        \caption{Sparse, truncation}
        \label{fig:xi_2000_sparse_trunc}
    \end{subfigure}
    \hspace{0.01\textwidth}
    \begin{subfigure}[b]{0.45\textwidth}
        \centering
        \includegraphics[width=\textwidth, keepaspectratio]{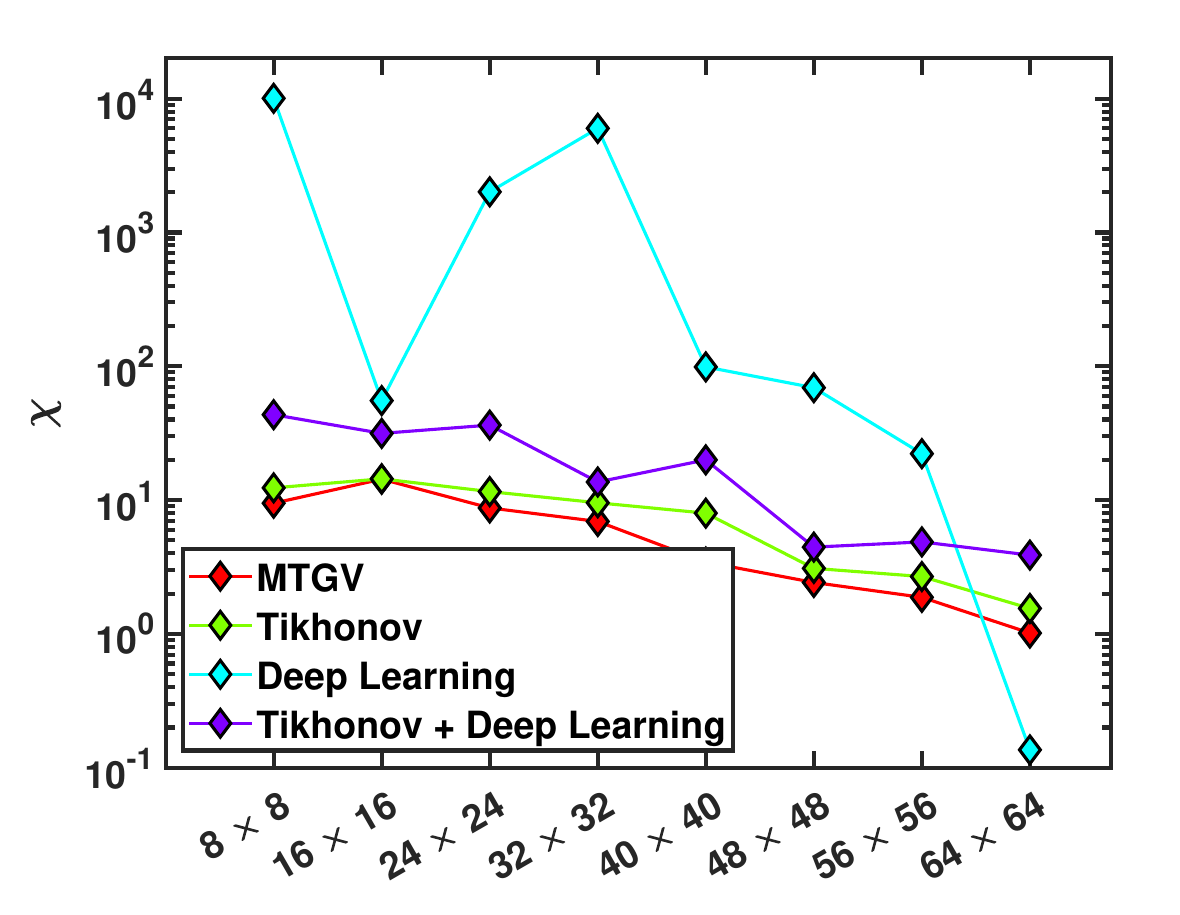}
        \caption{Smooth, truncation}
        \label{fig:xi_2000_smooth_trunc}
    \end{subfigure}
    \\
    \begin{subfigure}[b]{0.45\textwidth}
        \centering
        \includegraphics[width=\textwidth, keepaspectratio]{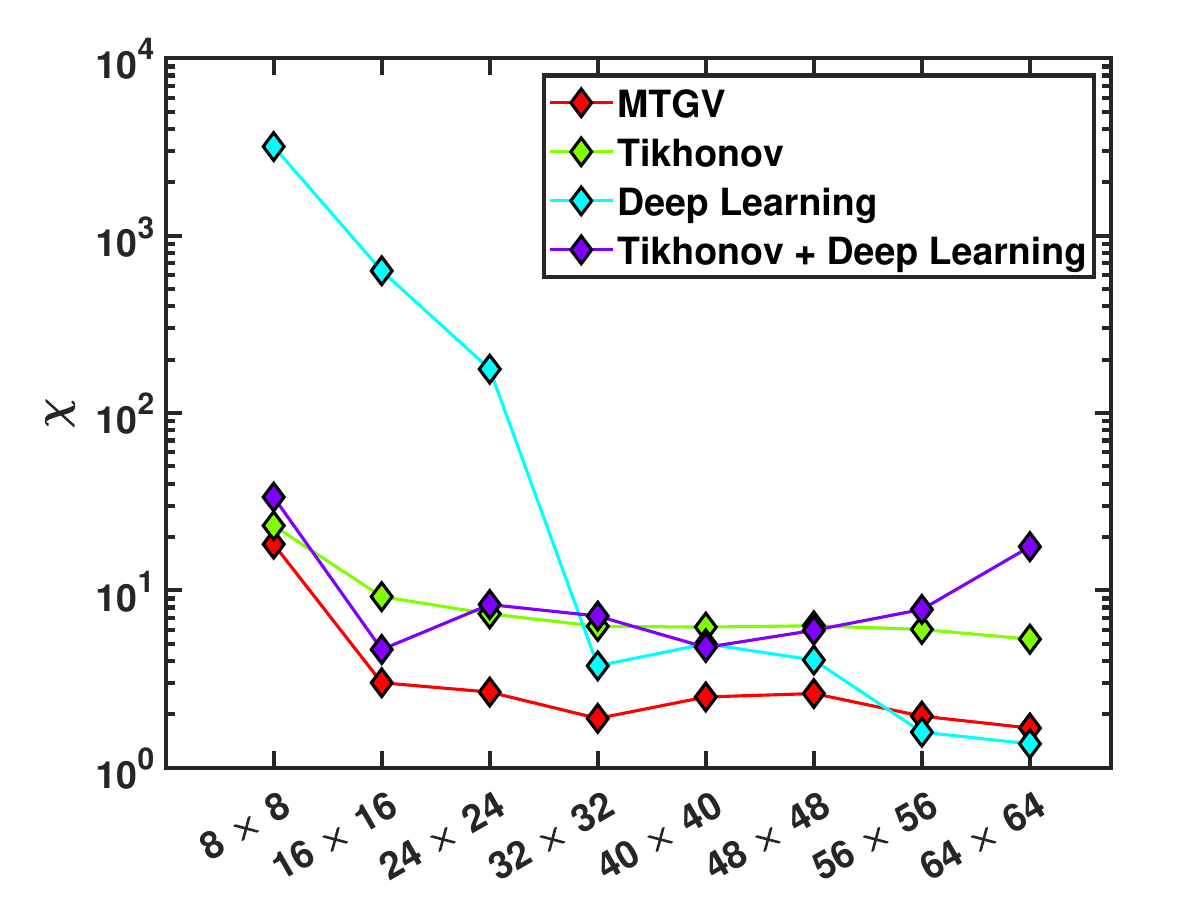}
        \caption{Sparse, truncation + random sampling}
        \label{fig:xi_2000_sparse_trunc_rs}
    \end{subfigure}
    \hspace{0.01\textwidth}
    \begin{subfigure}[b]{0.45\textwidth}
        \centering
        \includegraphics[width=\textwidth, keepaspectratio]{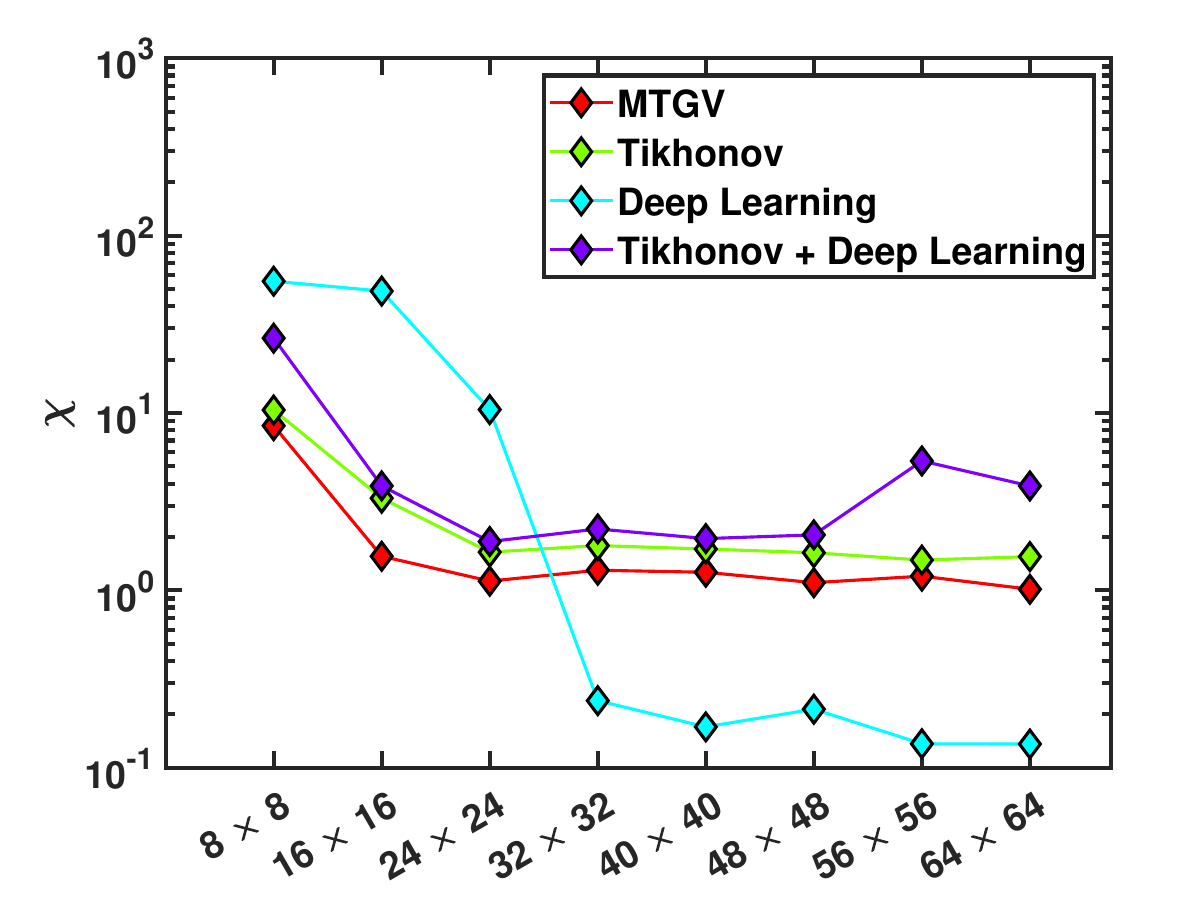}
        \caption{Smooth, truncation + random sampling}
        \label{fig:xi_2000_smooth_trunc_rs}
    \end{subfigure}
    \caption{$\chi$-score as defined through equation~\ref{eq:score_us} of the reconstructions obtained from artificial signals generated by the sparse~(left) and smooth~(right) $T_1$-$D$ distribution (figure~\ref{fig:dis_real_sparse} and~\ref{fig:dis_real_smooth}). The signal-to-noise ratio of the signal employed for inversion was 2000.}
    \label{fig:xi_2000}
\end{figure}

\begin{figure}[t]
    \centering
    \begin{subfigure}[b]{0.45\textwidth}
        \centering
        \includegraphics[width=\textwidth, keepaspectratio]{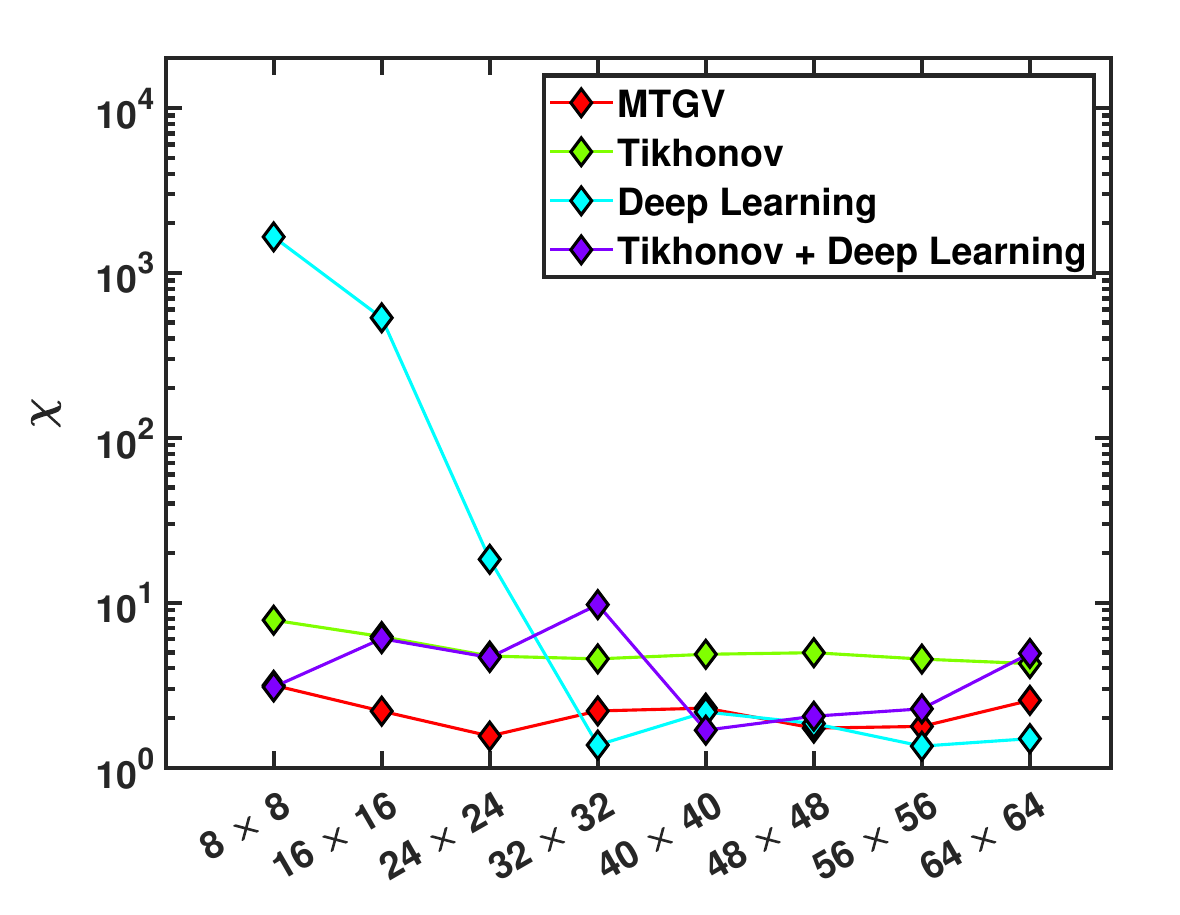}
        \caption{Random sampling}
        \label{fig:xi_2000_sm_sp_rs}
    \end{subfigure}
    \begin{subfigure}[b]{0.45\textwidth}
        \centering
        \includegraphics[width=\textwidth, keepaspectratio]{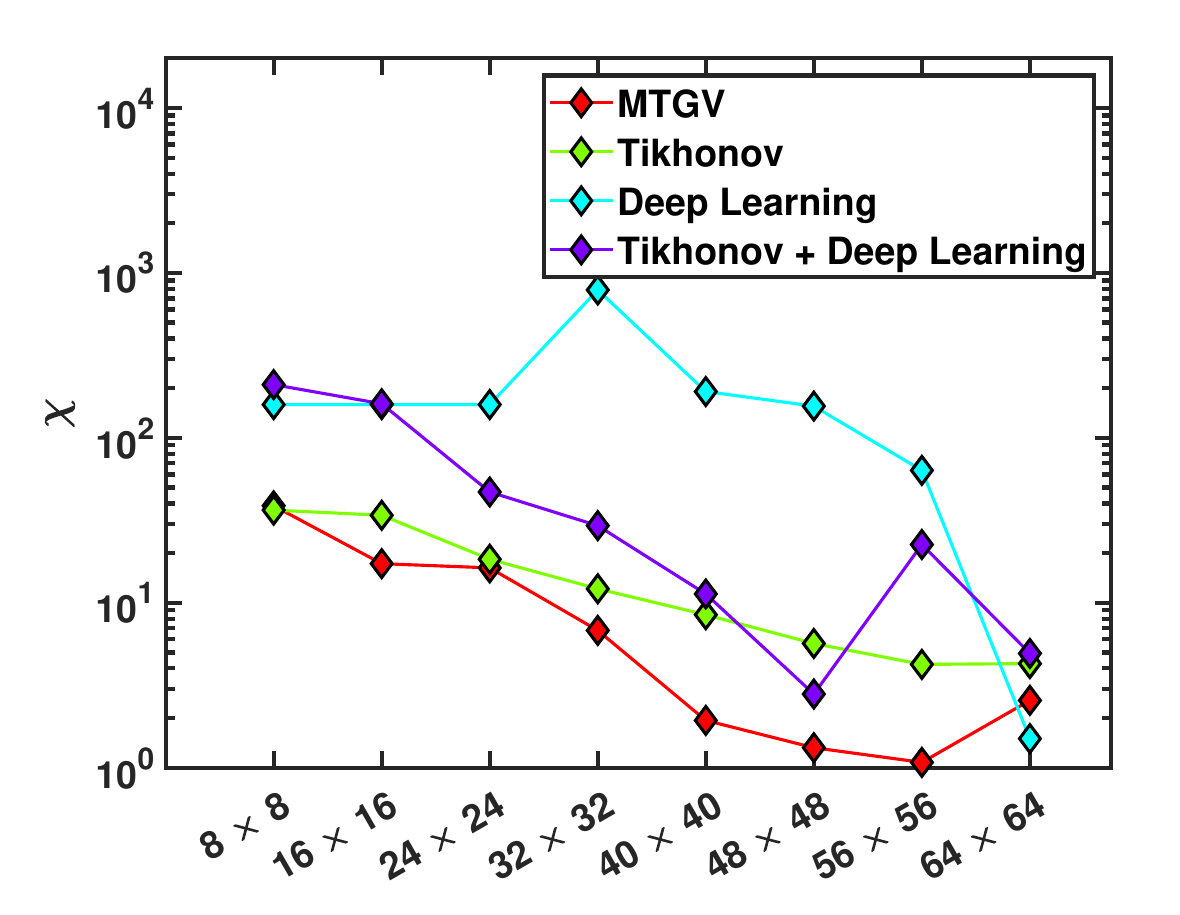}
        \caption{Truncation}
        \label{fig:xi_2000_sm_sp_trunc}
    \end{subfigure}
    \\
    \begin{subfigure}[b]{0.45\textwidth}
        \centering
        \includegraphics[width=\textwidth, keepaspectratio]{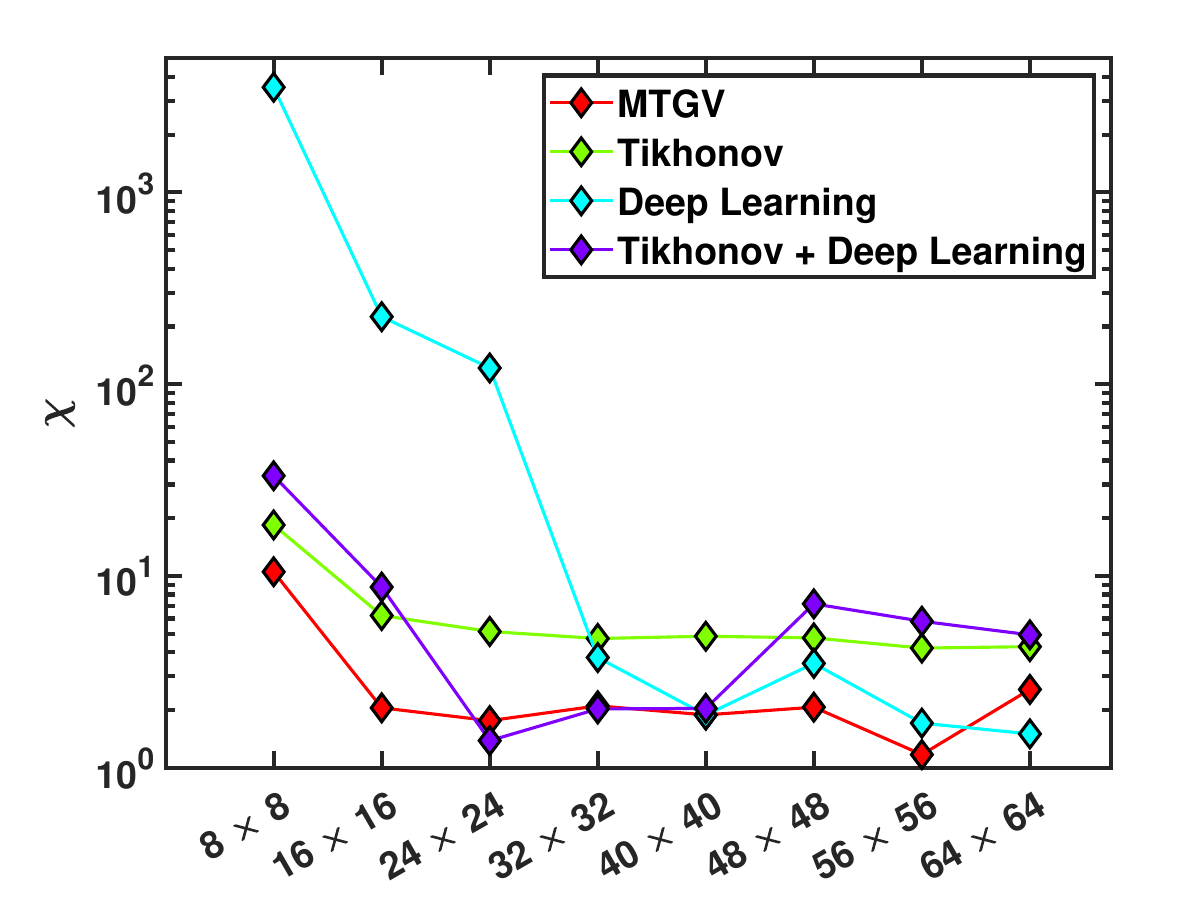}
        \caption{Truncation + random sampling}
        \label{fig:xi_2000_sm_sp_trunc_rs}
    \end{subfigure}
    \caption{$\chi$-score as defined through equation~\ref{eq:score_us} of the reconstructions obtained from artificial signals generated by the version of the $T_1$-$D$ distribution which contains smooth and sparse components (figure~\ref{fig:dis_real_sm_sp}). The signal-to-noise ratio of the signal employed for inversion was 2000.}
    \label{fig:xi_2000_sm_sp}
\end{figure}

\begin{figure}[t]
    \centering
    \begin{subfigure}[b]{0.45\textwidth}
        \centering
        \includegraphics[width=\textwidth, keepaspectratio]{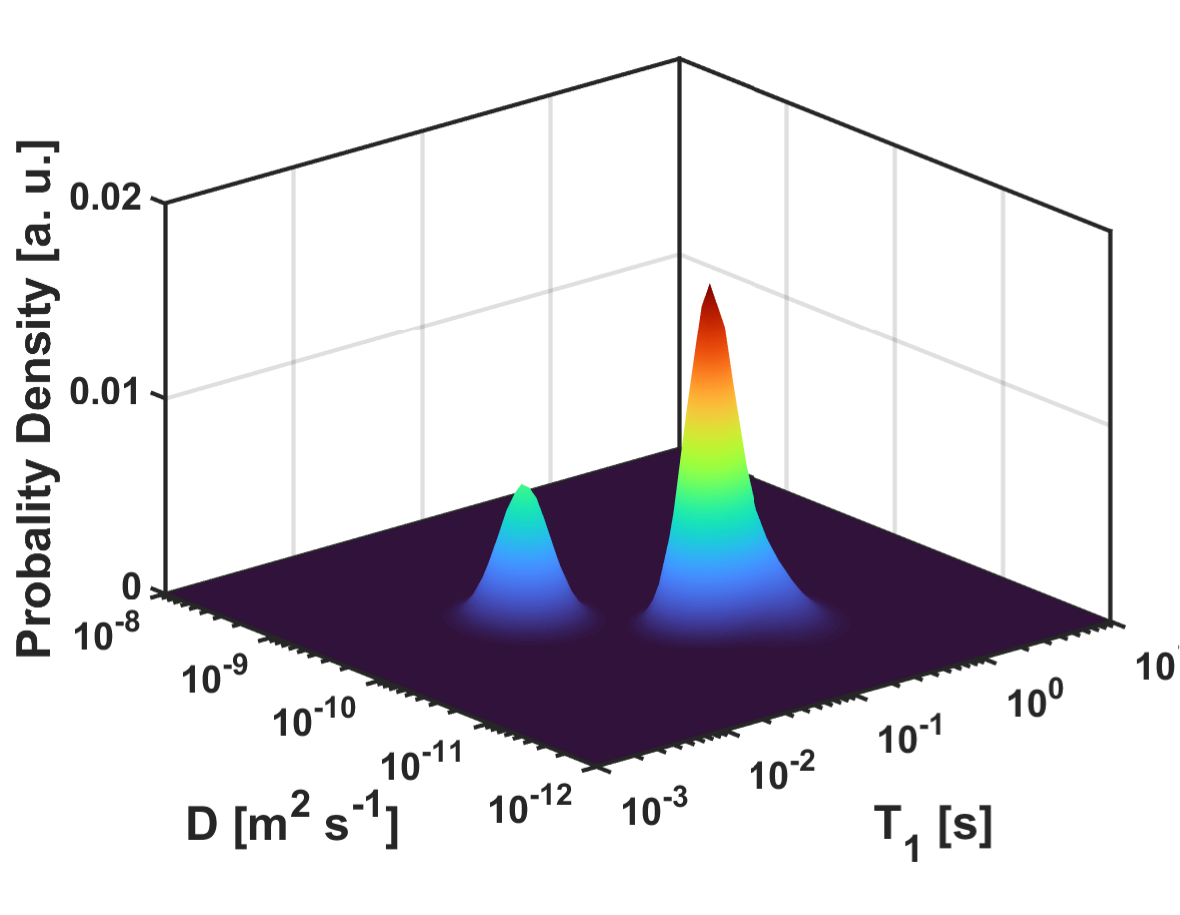}
        \caption{Real distribution}
        \label{fig:dis_real_smooth_copy}
    \end{subfigure}
    \hspace{0.02\textwidth}
    \begin{subfigure}[b]{0.45\textwidth}
        \centering
        \includegraphics[width=\textwidth, keepaspectratio]{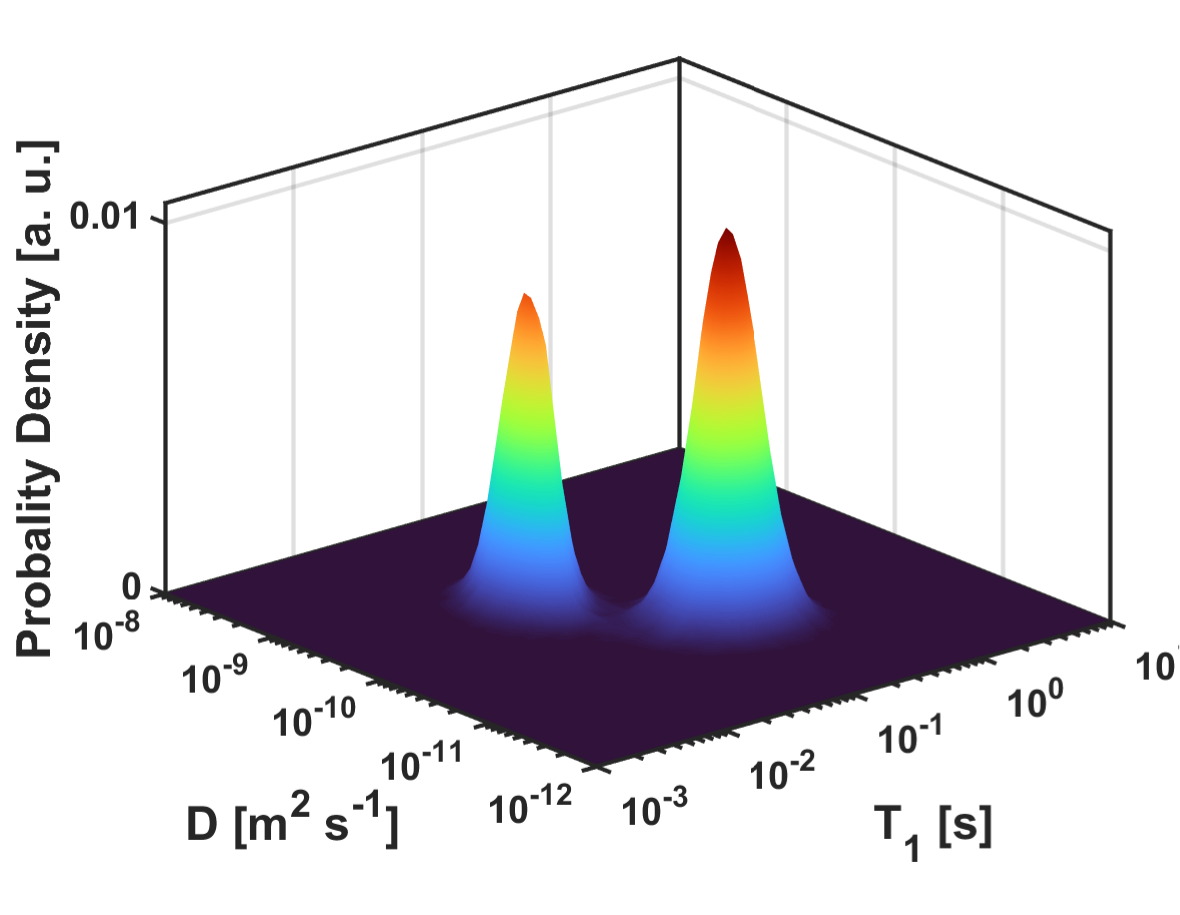}
        \caption{Deep learning}
        \label{fig:dis_smooth_deep}
    \end{subfigure}
    \\
    \begin{subfigure}[b]{0.45\textwidth}
        \centering
        \includegraphics[width=\textwidth, keepaspectratio]{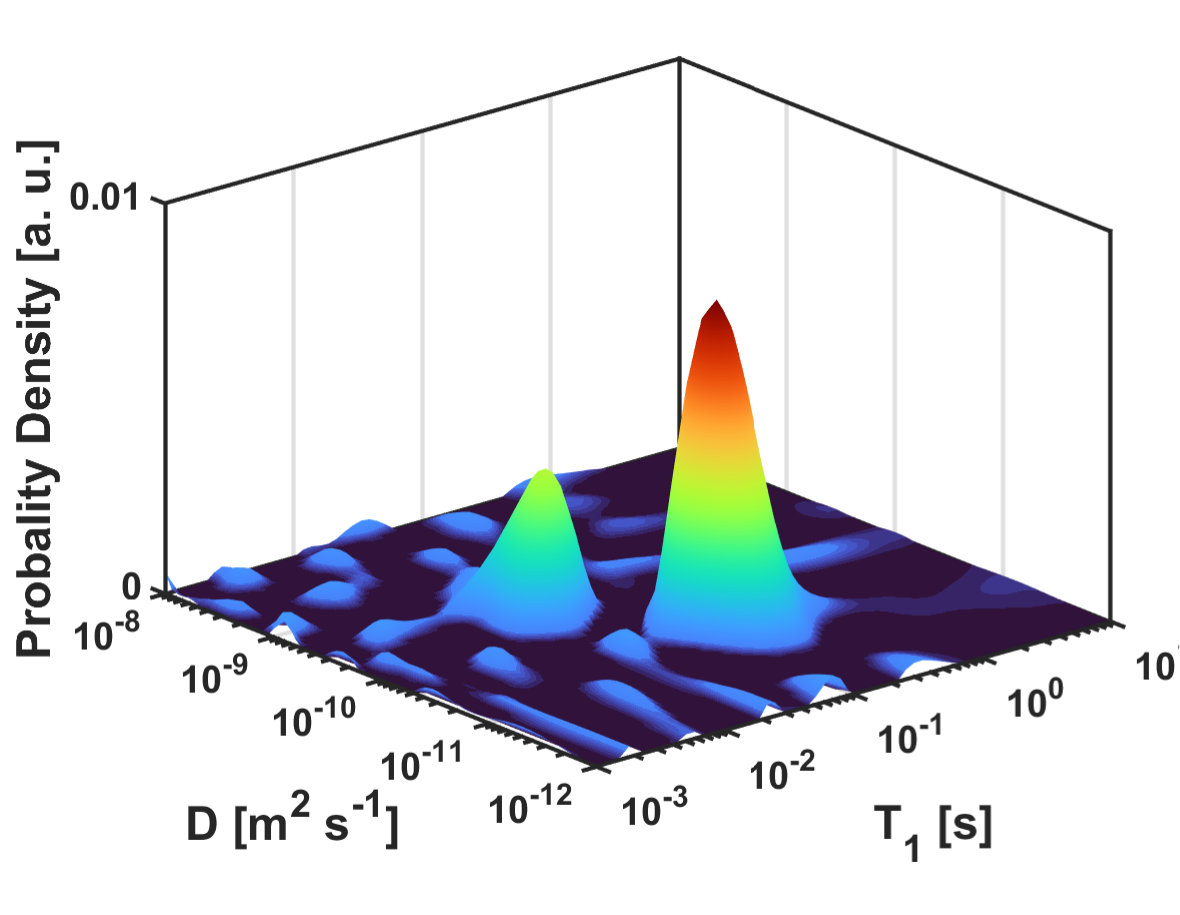}
        \caption{Tikhonov}
        \label{fig:dis_smooth_tik}
    \end{subfigure}
    \hspace{0.02\textwidth}
    \begin{subfigure}[b]{0.45\textwidth}
        \centering
        \includegraphics[width=\textwidth, keepaspectratio]{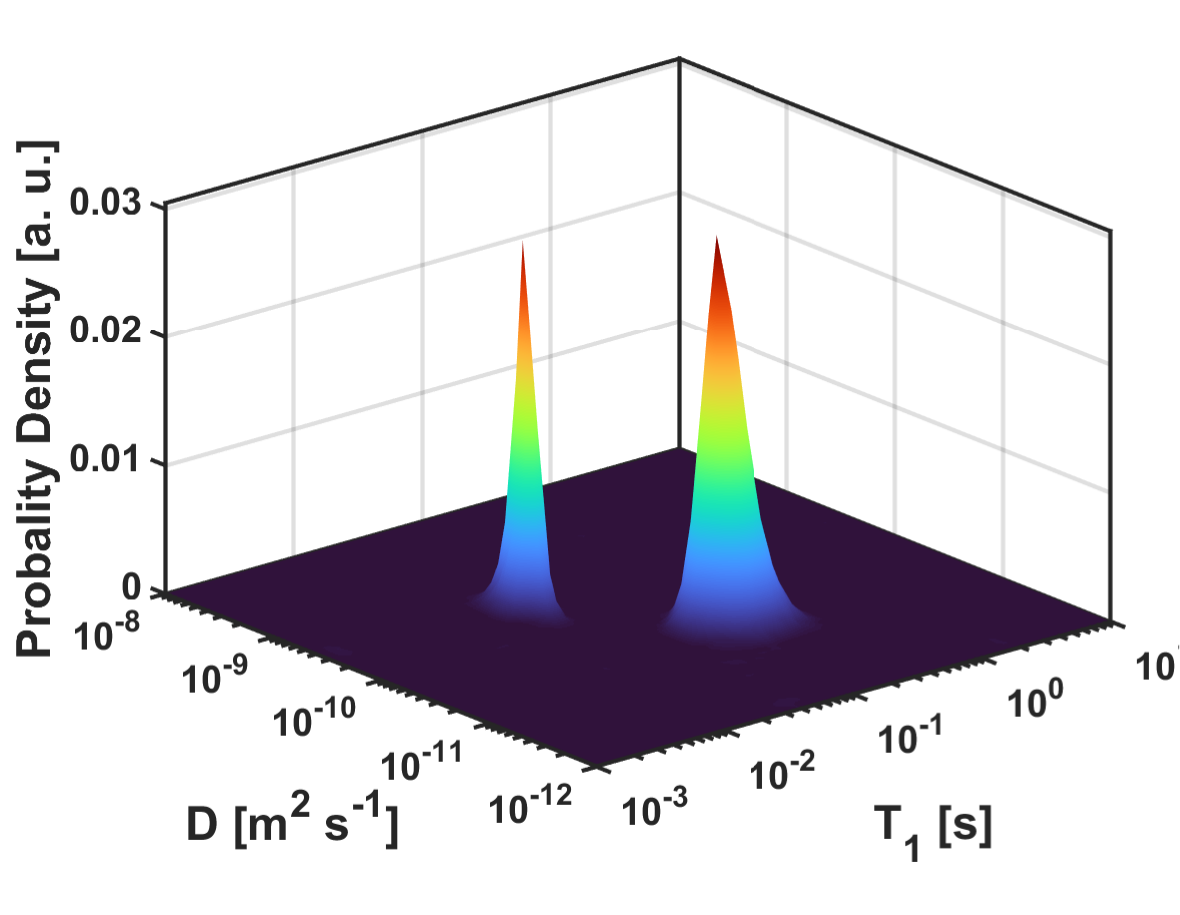}
        \caption{Tikhonov + deep learning}
        \label{fig:dis_smooth_tik_deep}
    \end{subfigure}
    \caption{Reconstructed distributions (excluding the result from MTGV) obtained from artificial signals generated by the version of the $T_1$-$D$ distribution which only contains smooth components (figure~\ref{fig:dis_real_smooth}). The signal-to-noise ratio of the signal employed for inversion was 2000. For the reason of a simpler comparison, the real distribution was re-scaled and the updated plot is shown in figure~\ref{fig:dis_real_smooth_copy}.}
    \label{fig:dis_rec}
\end{figure}

\begin{figure}[t!]
    \centering
    \begin{subfigure}[b]{0.45\textwidth}
        \centering
        \includegraphics[width=\textwidth, keepaspectratio]{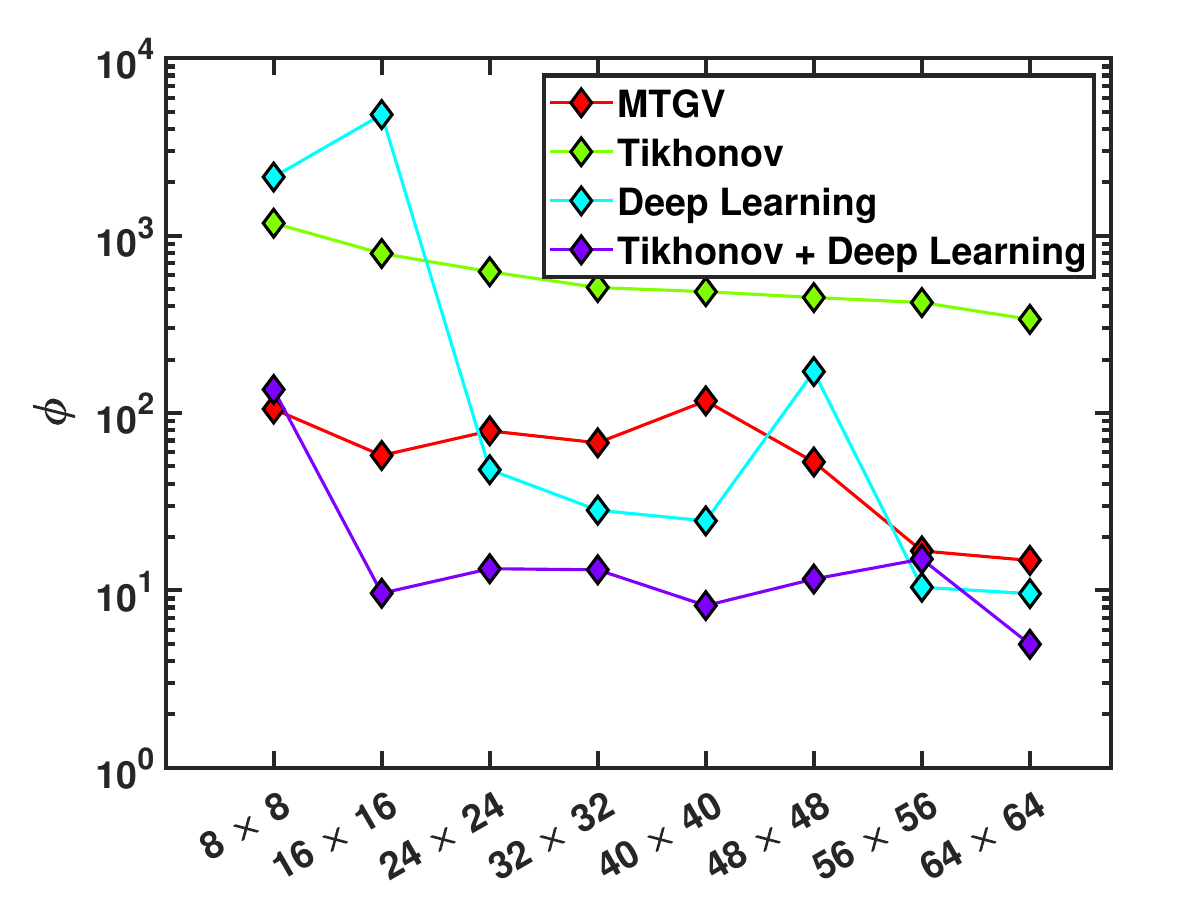}
        \caption{Sparse, random sampling}
        \label{fig:phi_2000_sparse_rs}
    \end{subfigure}
    \hspace{0.01\textwidth}
    \begin{subfigure}[b]{0.45\textwidth}
        \centering
        \includegraphics[width=\textwidth, keepaspectratio]{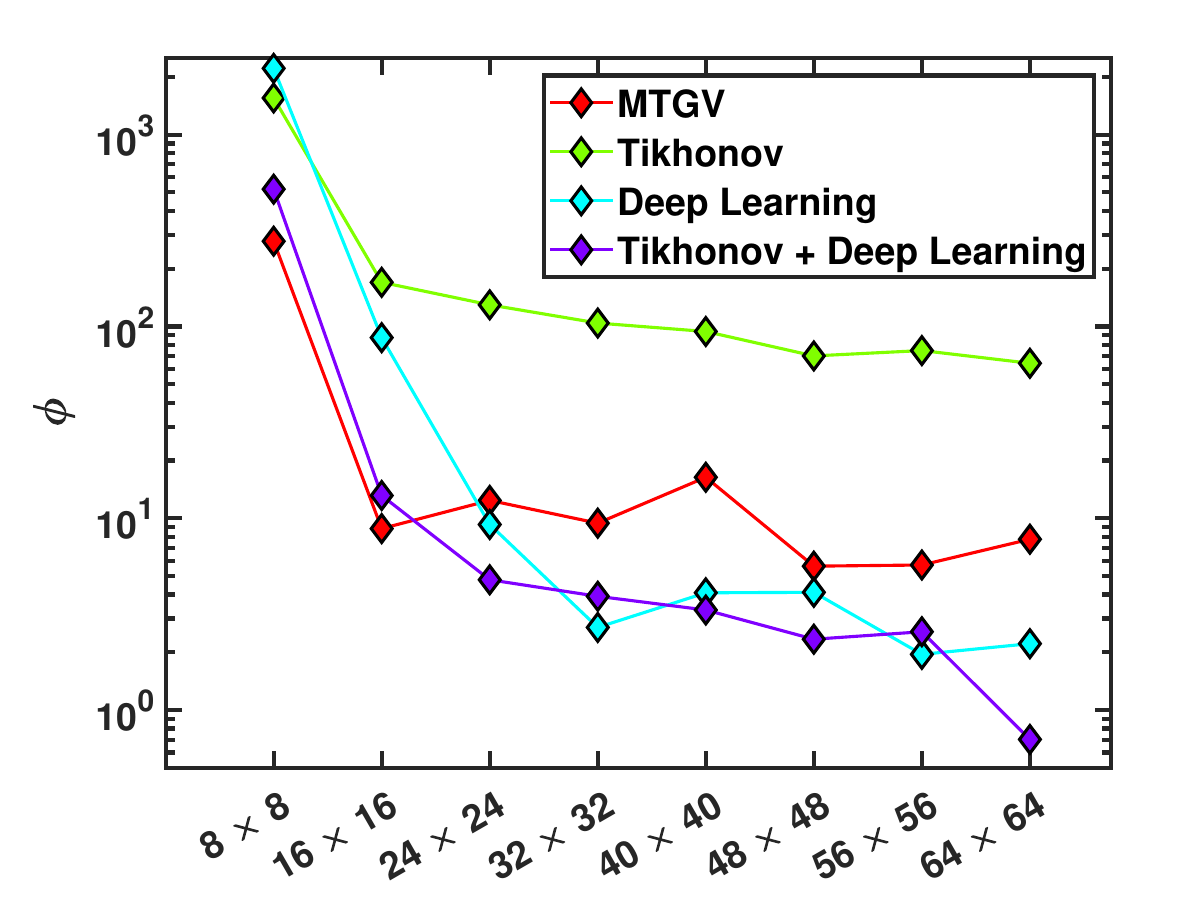}
        \caption{Smooth, random sampling}
        \label{fig:phi_2000_smooth_rs}
    \end{subfigure}
    \\
    \begin{subfigure}[b]{0.45\textwidth}
        \centering
        \includegraphics[width=\textwidth, keepaspectratio]{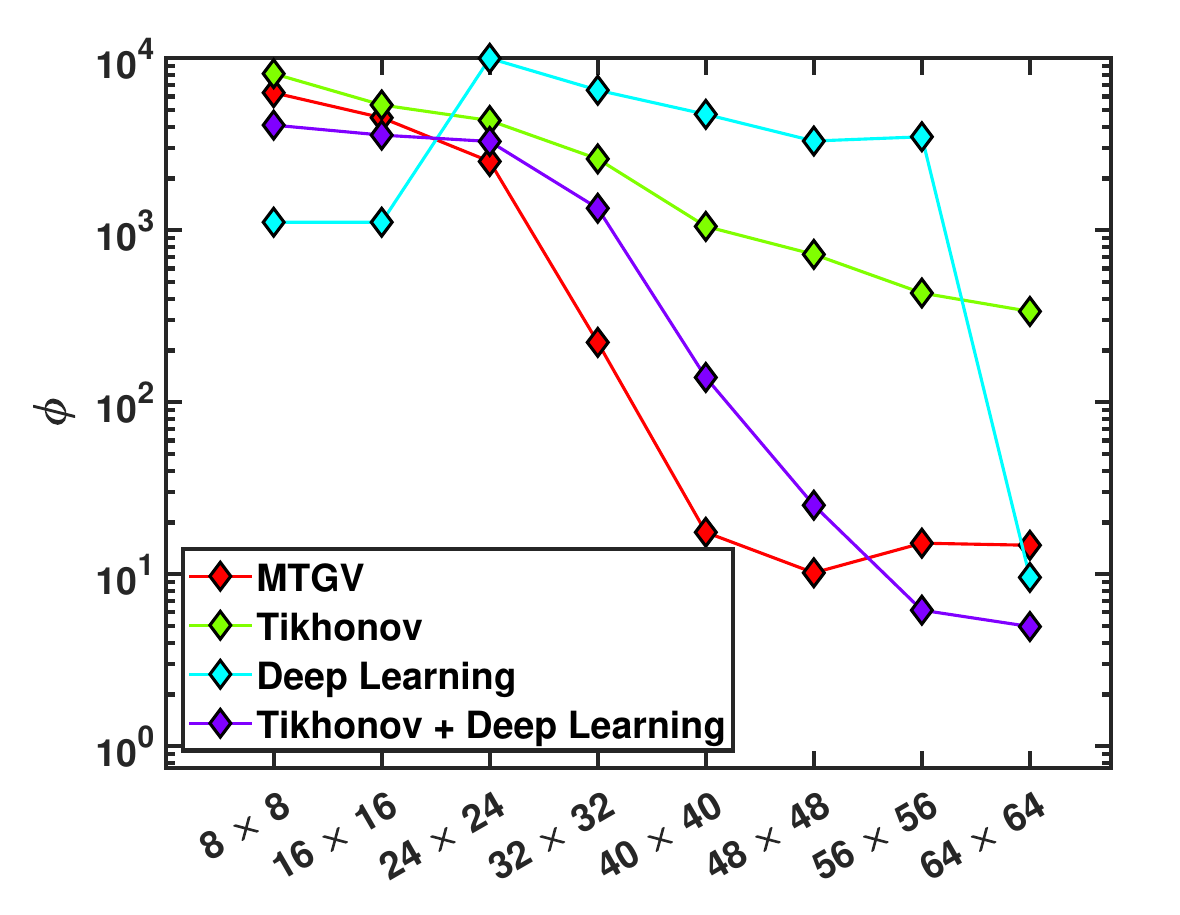}
        \caption{Sparse, truncation}
        \label{fig:phi_2000_sparse_trunc}
    \end{subfigure}
    \hspace{0.01\textwidth}
    \begin{subfigure}[b]{0.45\textwidth}
        \centering
        \includegraphics[width=\textwidth, keepaspectratio]{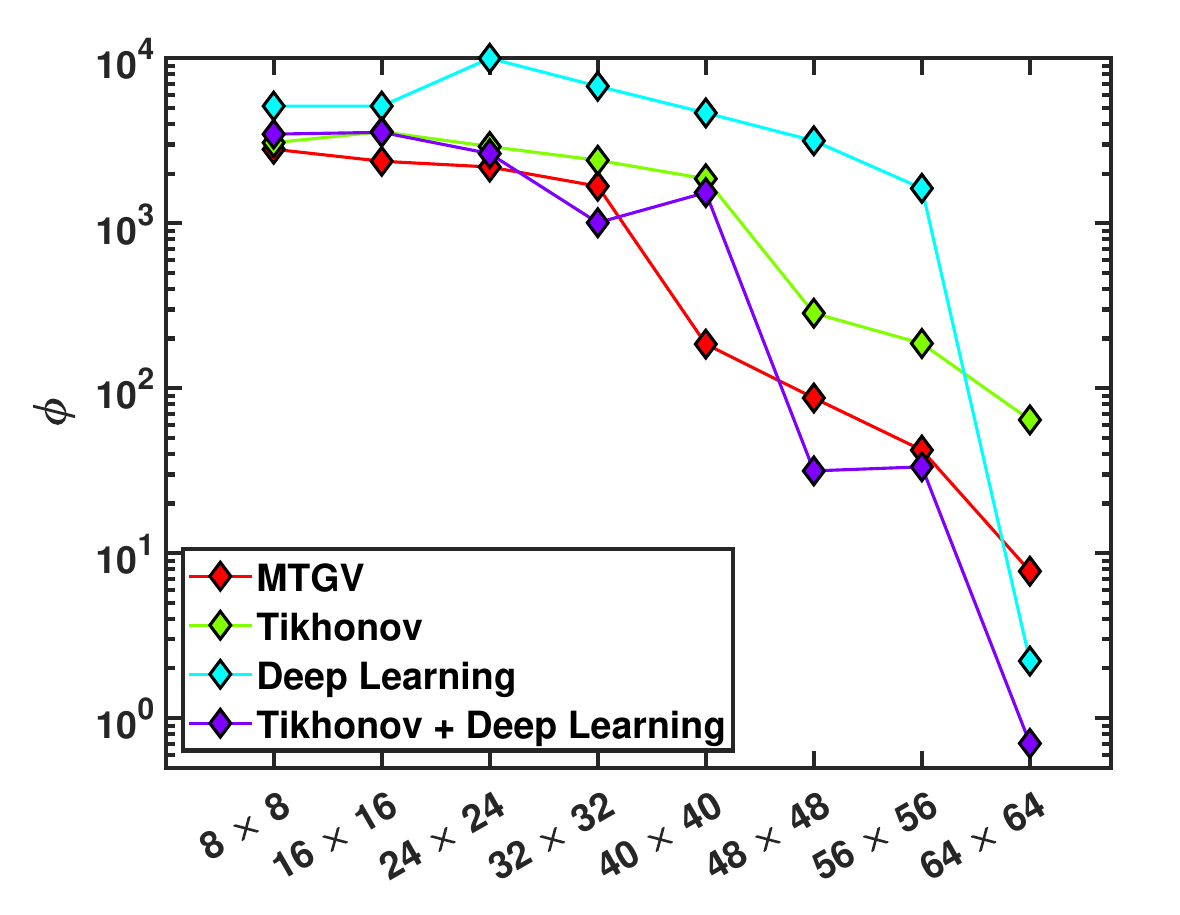}
        \caption{Smooth, truncation}
        \label{fig:phi_2000_smooth_trunc}
    \end{subfigure}
    \\
    \begin{subfigure}[b]{0.45\textwidth}
        \centering
        \includegraphics[width=\textwidth, keepaspectratio]{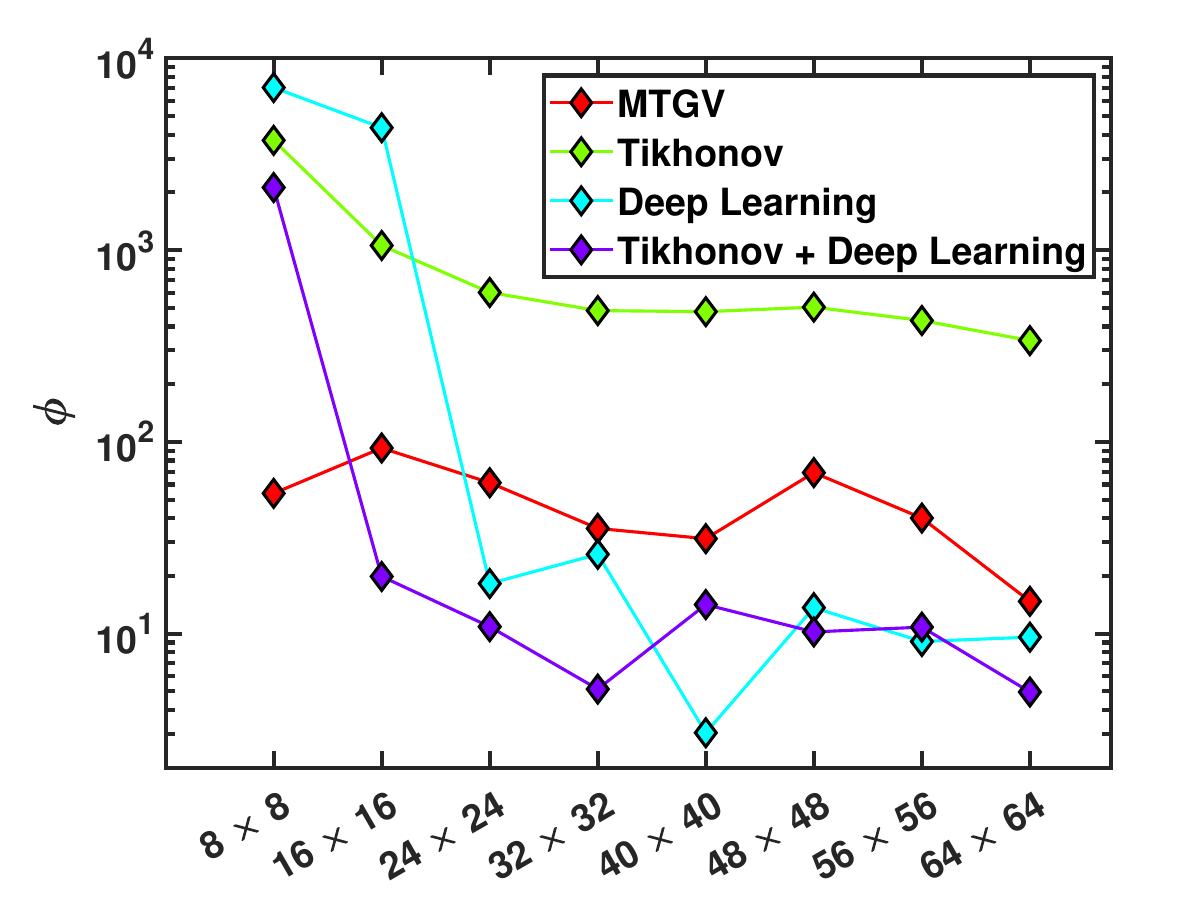}
        \caption{Sparse, truncation + random sampling}
        \label{fig:phi_2000_sparse_trunc_rs}
    \end{subfigure}
    \hspace{0.01\textwidth}
    \begin{subfigure}[b]{0.45\textwidth}
        \centering
        \includegraphics[width=\textwidth, keepaspectratio]{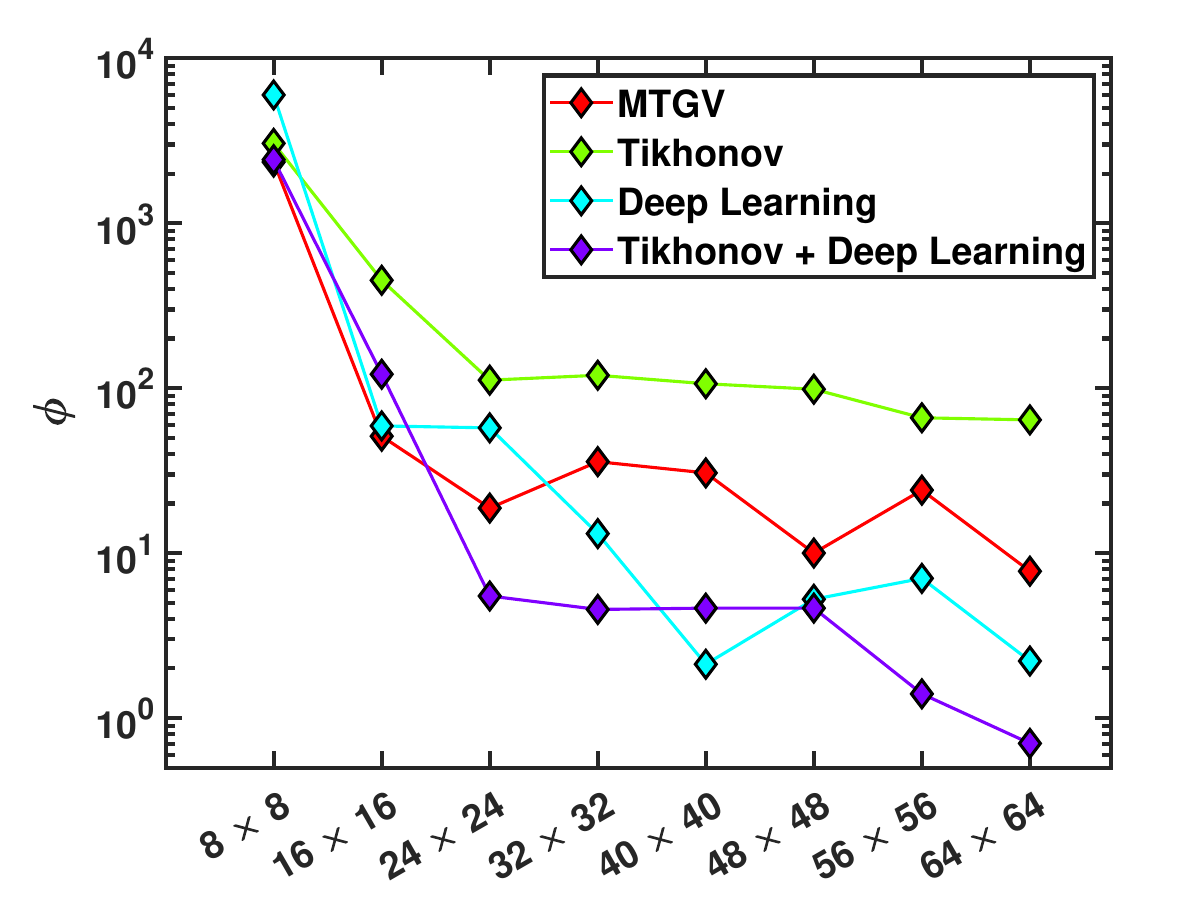}
        \caption{Smooth, truncation + random sampling}
        \label{fig:phi_2000_smooth_trunc_rs}
    \end{subfigure}
    \caption{$\phi$-score as defined through equation~\ref{eq:score_us_art} of the reconstructions obtained from artificial signals generated by the sparse~(left) and smooth~(right) $T_1$-$D$ distribution (figure~\ref{fig:dis_real_sparse} and~\ref{fig:dis_real_smooth}). The signal-to-noise ratio of the signal employed for inversion was 2000.}
    \label{fig:phi_2000}
\end{figure}

\begin{figure}[t]
    \centering
    \begin{subfigure}[b]{0.45\textwidth}
        \centering
        \includegraphics[width=\textwidth, keepaspectratio]{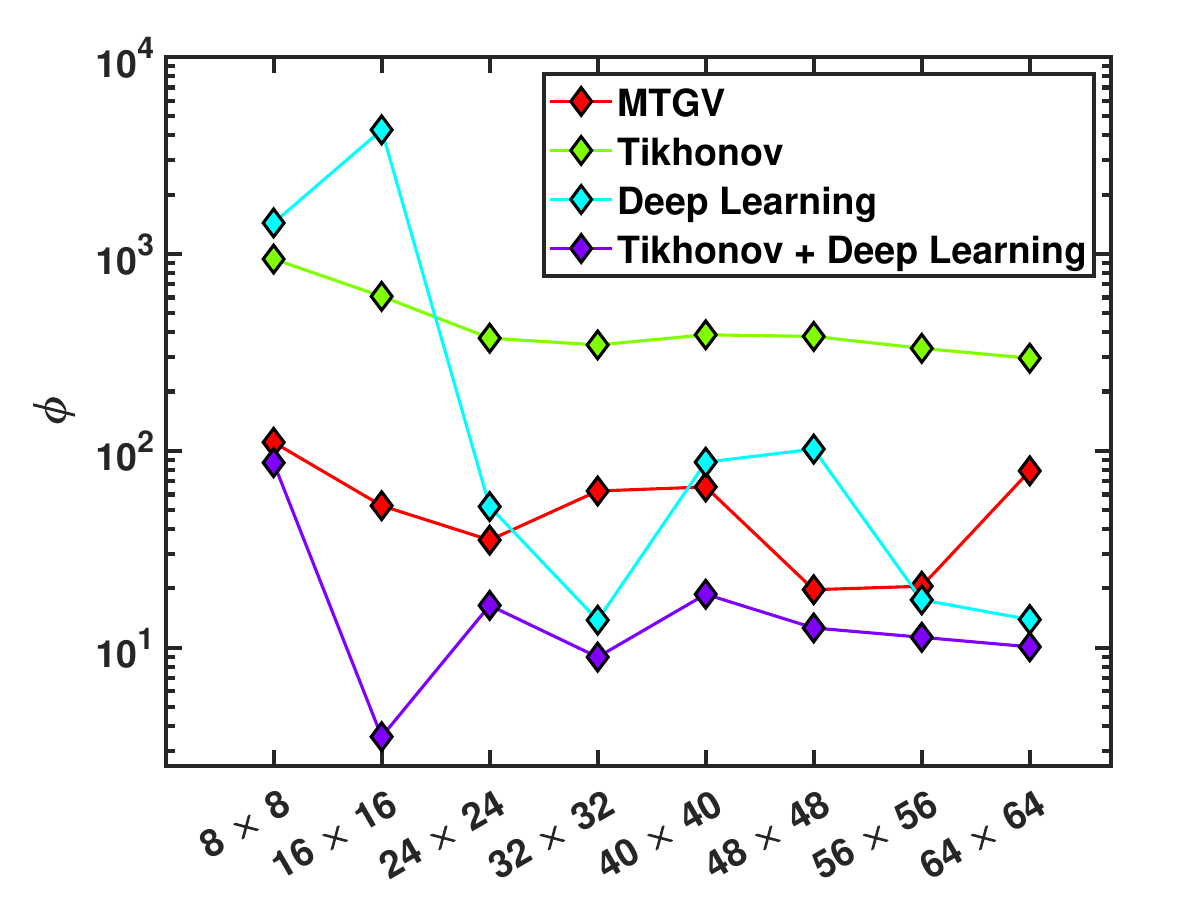}
        \caption{Random sampling}
        \label{fig:phi_2000_sm_sp_rs}
    \end{subfigure}
    \begin{subfigure}[b]{0.45\textwidth}
        \centering
        \includegraphics[width=\textwidth, keepaspectratio]{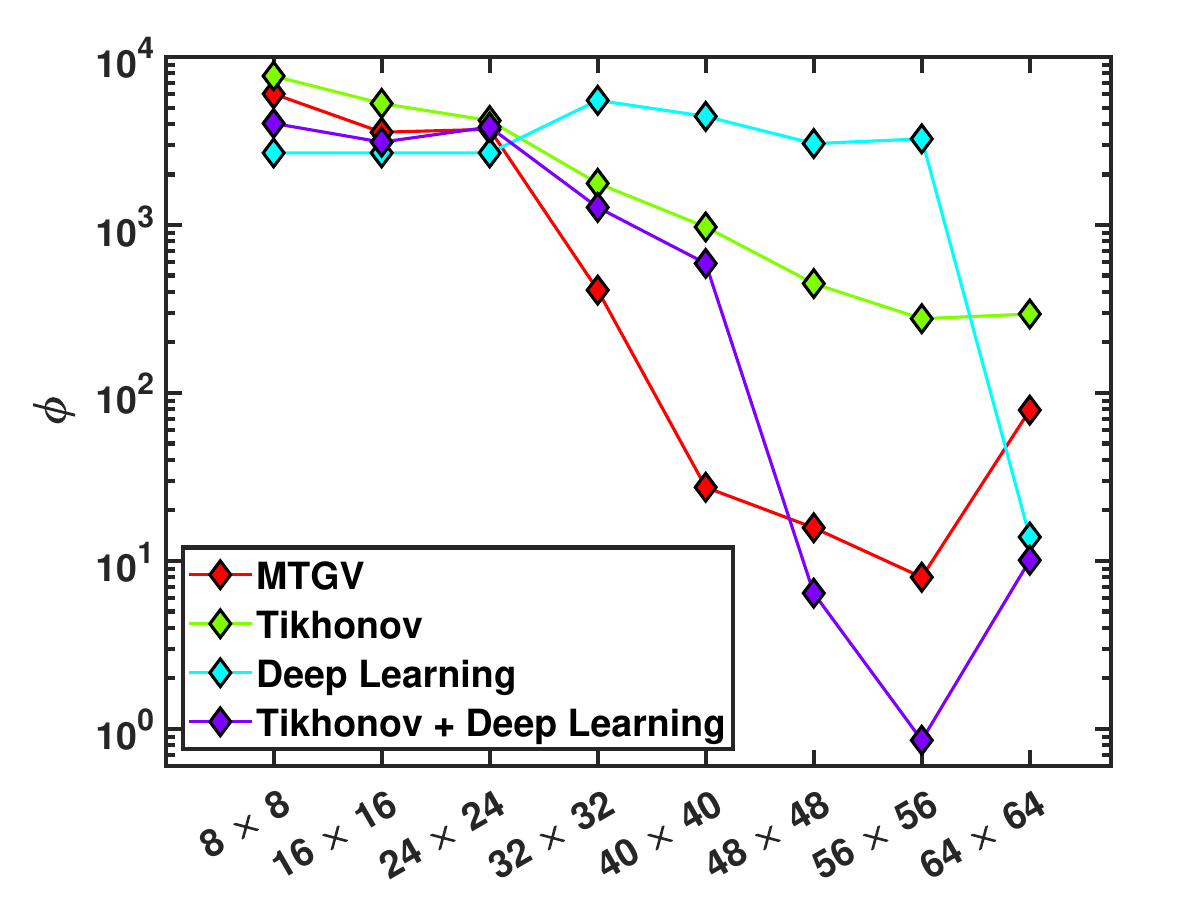}
        \caption{Truncation}
        \label{fig:phi_2000_sm_sp_trunc}
    \end{subfigure}
    \\
    \begin{subfigure}[b]{0.45\textwidth}
        \centering
        \includegraphics[width=\textwidth, keepaspectratio]{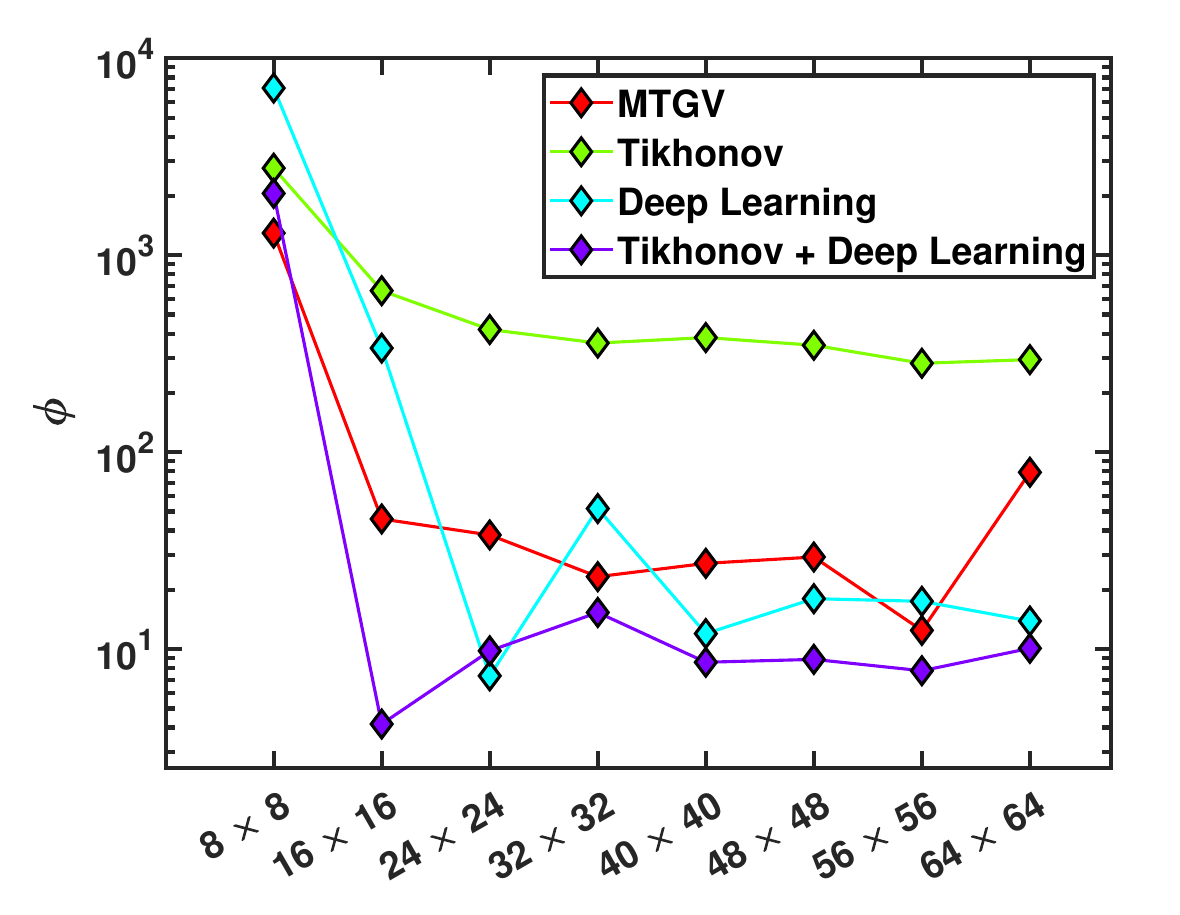}
        \caption{Truncation + random sampling}
        \label{fig:phi_2000_sm_sp_trunc_rs}
    \end{subfigure}
    \caption{$\phi$-score as defined through equation~\ref{eq:score_us_art} of the reconstructions obtained from artificial signals generated by the version of the $T_1$-$D$ distribution which contains smooth and sparse components (figure~\ref{fig:dis_real_sm_sp}). The signal-to-noise ratio of the signal employed for inversion was 2000.}
    \label{fig:phi_2000_sm_sp}
\end{figure}

The scores obtained from equation~\ref{eq:score_us} plotted against the number of points in the associated sub-sampling subset are shown in figure~\ref{fig:xi_2000}. For the reason of brevity only the results from signals with a signal-to-noise ratio of 2000 are given at this point, but the remaining figures including the signal-to-noise ratios 200 and 20 can be found in previous work of Beckmann \textit{et al.}.\supercite{Beckmann_phd} However, comparable results in-between different signal-to-noise ratios could be identified with most outliers for the combined method of Tikhonov and deep learning, but overall trends remain close to constant independent of the noise level.  
From figure~\ref{fig:xi_2000} and~\ref{fig:xi_2000_sm_sp} it becomes evident, that for the fully sampled signal deep learning outperforms all other inversion methods independent of the sparsity and smoothness of the original distribution, followed by MTGV which unambiguously shows the second best performance of all reconstruction methods. For distributions containing at least one sparse components, the difference between MTGV and deep learning is significant but not enormous (figure~\ref{fig:xi_2000} and~\ref{fig:xi_2000_sm_sp}), whereas for the exclusively smooth distribution the difference between deep learning and all other inversion methods is one order of magnitude or more. Moving to sub-sampled signals stark differences between the employed sub-sampling methods are revealed. Here, the most considerable effect is found for the combination of deep learning and signal truncation (figure~\ref{fig:xi_2000_sparse_trunc},~\ref{fig:xi_2000_smooth_trunc} and~\ref{fig:xi_2000_sm_sp_trunc}). Comparing, the deep learning results of the fully sampled signals with the first truncated subset consisting of 56~$\times$~56 data points, the $\chi$-score approximately increases two order of magnitude post signal truncation. This effect continues to grow if the signal is cut off earlier, but only less clear trends can be identified. For instance, independent of the sparsity and smoothness of the original distribution, the 32 $\times$ 32 truncated subsets reaches a ranking of close to $10^3$ or even higher, which decreases again of roughly one order of magnitude or more if the 16 $\times$ 16 truncated data set is considered. Focusing on the combination of truncation and the remaining inversion methods, it becomes evident that with the exception of some outliers more significant truncation results in higher rankings with the maximum reached if the 8 $\times$ 8 or 16 $\times$ 16 subsets are employed, but overall differences are small compared to the combined approach of deep learning and truncation. In general, this finding holds true for all inversion methods except deep learning and is further independent on the level of sparsity and smoothness of the original distribution. 
Altogether, signal truncation shows the best results in combination with MTGV followed by Tikhonov and its combination with deep learning and undoubtedly the worst performance if deep learning is employed. Moving from truncation to random sampling (figure~\ref{fig:xi_2000_sparse_rs},~\ref{fig:xi_2000_smooth_rs} and~\ref{fig:xi_2000_sm_sp_rs}), it becomes evident that deep learning exhibits the best performance in the vast majority of instances if at least 32 $\times$ 32 or more data points are sampled. In comparison, the combination of MTGV and Tikhonov with random sampling results in an overall close to constant performance for both regularization method with MTGV clearly outperforming Tikhonov. This finding is independent of the level of sub-sampling with some exceptions for the smooth distribution as shown in figure~\ref{fig:xi_2000_smooth_rs}. In contrast, the performance of the combined approach of deep learning and Tikhonov does not allow for the identification of a clear trend, instead $\chi$-scores show fluctuations with commonly worse performance of the fully sampled signal compared to significant random sub-sampling. Focusing on the combination of truncation and random sampling as a sub-sampling method (figure~\ref{fig:xi_2000_sparse_trunc_rs},~\ref{fig:xi_2000_smooth_trunc_rs} and~\ref{fig:xi_2000_sm_sp_trunc_rs}), similar trends to random sampling alone can be identified. For MTGV and Tikhonov again close to constant trends with MTGV ahead in performance are revealed with the exception of a considerable performance loss in the case of the 8 $\times$ 8 subset. For the combined approach of Tikhonov and deep learning, the rankings reveal significant fluctuations again with the peculiarity that the 8 $\times$ 8 subset but also the fully sampled signal show the highest rankings compared to majority of sub-samples in-between. In general, the fluctuating behaviour and poor performance of the combined approach of Tikhonov and deep learning is something unexpected. For instance, the fact that Tikhonov and deep learning in combination show the worst performance for the fully sampled signal across all other tested methods as well as independent of the level of sparsity and smoothness of the original distribution is not \hl{straightforward} to comprehend. Based on the principles \hl{outlined} in section~\ref{sec:US_meth}, the neural network is assigned to act as a filter removing artefacts from the reconstructions obtained via Tikhonov regularization. Hence, it would be expected that this combined approach outperforms Tikhonov regularization alone at least. To gain deeper insights for the reason of those counter-intuitive results, the reconstructed distributions of the original smooth $T_1$-$D$ distribution are plotted for all employed inversion methods with the exception of MTGV and are shown in figure~\ref{fig:dis_rec}. To facilitate the comparison between the inversion results and the real distribution, figure~\ref{fig:dis_real_smooth} was re-scaled and subsequently re-plotted in figure~\ref{fig:dis_rec}. 
A comparison of the inversion results with the real distribution shows that for all inversion methods tested a high-quality reconstruction was obtained and that even for the combined approach of Tikhonov and deep learning which received the highest $\chi$-score, the reconstruction can be considered as a good approximation and it would be fit for purpose for many applications. In contrast, the reconstruction received from Tikhonov regularization contains a vast number of artefacts peaks, which makes an unambiguous interpretation difficult. This is a major concern from an experimental point of view, because in the most instances the ground truth distribution is not known \hl{prior to} the experiment, which renders the occurrence of additional artefact peaks particularly problematic, especially if the physics of the investigated system is not well understood and hence, small but significant contributions from surface interactions or exchange cannot be ruled out. Consequently, artefact peaks are a major concern for the unambiguous interpretation of experimental results. From this point of view, it can be argued that figure~\ref{fig:dis_smooth_tik_deep} is actually the better reconstruction results compared to the distribution obtained through Tikhonov regularization (figure~\ref{fig:dis_smooth_tik}). This poses the question for the reason of Tikhonov regularization receiving a better ranking than the combined approach of Tikhonov and deep learning. In this context, the rankings and the inversion results indicate that the narrow and sparse peak form in figure~\ref{fig:dis_smooth_tik_deep} is more heavily penalized than the artefact peaks in figure~\ref{fig:dis_smooth_tik}. Therefore, the definition of an additional penalty function which penalizes reconstruction artefacts very heavily is expected to be beneficial for an unambiguous interpretation of experimental results especially if the ground truth distribution is unknown and the physics of the investigated system is only poorly understood. For this reason, the reconstructed results are re-ranked using equation~\ref{eq:score_us_art}. The obtained $\phi$-rankings are shown in figure~\ref{fig:phi_2000} and~\ref{fig:phi_2000_sm_sp}. Using equation~\ref{eq:score_us_art} instead of equation~\ref{eq:score_us}, impacts the relative positioning of the rankings of the employed inversion methods in a very notable manner. For the fully sampled signal, the combination of Tikhonov and deep learning exhibits the best performance independent of the level of sparsity and smoothness in the original distribution. This means the method performing worst based on the $\chi$-score metric is ranking best if equation~\ref{eq:score_us_art} is employed as an evaluation method instead. Considering sub-sampling across all employed sub-sampling methods, combining Tikhonov and deep learning persists to be the best inversion method overall, outperforming the remaining reconstructions techniques in a clear majority of instances. 
Focusing on random sampling (figure~\ref{fig:phi_2000_sparse_rs},~\ref{fig:phi_2000_smooth_rs} and~\ref{fig:phi_2000_sm_sp_rs}), Tikhonov regularization displays the worst performance for nearly all data sets tested with only deep learning showing even higher rankings if maximum or close-to-maximum sub-sampling is used. However, a considerable increase of the $\phi$-score close to the maximum sub-sampling level can be observed for all inversion methods independent of original distribution. In addition, a comparison of MTGV and deep learning reveals less clear trends with MTGV outperforming deep learning in some instances and vice versa in others with the overall number of rankings slightly in favour of deep learning over MTGV. Moving from random sampling to the combination of truncation and random sampling (figure~\ref{fig:phi_2000_sparse_trunc_rs},~\ref{fig:phi_2000_smooth_trunc_rs} and~\ref{fig:phi_2000_sm_sp_trunc_rs}), comparable trends to random sampling alone prevail. In the case of truncation (figure~\ref{fig:phi_2000_sparse_trunc},~\ref{fig:phi_2000_smooth_trunc} and~\ref{fig:phi_2000_sm_sp_trunc}), a similar behaviour as already identified for equation~\ref{eq:score_us} can be observed. Compared to the fully sampled signal, truncation leads to significant higher $\phi$-scores usually increasing with growing truncation threshold (with the 8 $\times$ 8 subset coinciding with the highest truncation threshold). This trend is the most pronounced for deep learning but very significant effects persist for the remaining inversion methods, which are even beyond the observations made for equation~\ref{eq:score_us}. In general, it is further notable, that Tikhonov regularization aside,  equation~\ref{eq:score_us_art} shows a considerably more erratic behaviour as it is the case for  equation~\ref{eq:score_us}.\\

\noindent Firstly, focusing on the $\chi$-score results, the finding of deep learning outperforming all other inversion methods is consistent with the observations made in earlier work, highlighting the inversion capabilities of thoroughly trained neural networks. The sharp drop in network performance if truncation is employed could stem from insufficient network training on not-fully decayed signals and noise effects originating from interpolation of the truncated signal to fit the input size of the neural network. However, it is unlikely that this is actually the case here. During network training a very significant number of not-fully decayed signals is included in the training data which rules out the possibility that the observed performance drop after truncation is merely originating from poor network training. Further, the fact that truncation results in a sharp drop in network performance even at the lowest sub-sampling levels (coinciding with the 56 $\times$ 56 subset), which is not seen for random sampling and its combination with truncation, indicates that interpolation cannot account for the poor performance of truncated signals. 
Consequently, those findings provide evidence that the neural network under test distinguishes between distributions due to differences in the decay behaviour of the input signals and hence, a good reconstruction can only be achieved, if the full dynamic range of an input signal is sampled. This agrees with the findings obtained for the remaining inversion methods which also display worse performance for truncated and hence not-fully decayed signals, although the observed effect is not as significant as it is for deep learning. Overall, those observations highlight the importance of dynamic range on inversion performance and show that the best reconstruction can only be obtained if the full dynamic range of the to-be-inverted signal is covered. Hence, truncation only performs poorly compared to the other sub-sampling methods employed in this publication and should not be used for sub-sampling if random sampling is available. Moving to random sampling, the close-to-constant performance of MTGV and Tikhonov regularization, which is independent of the degree of sub-sampling, agrees with the dynamic range argument made for truncation and further highlights the importance to cover the full dynamic range of the signal decay in the subset employed for inversion. The significance of high dynamic range is also supported by the inversion results obtained from deep learning and randomly sampled subsets. The sharp drop in network performance starting at sub-sets of the size 24 $\times$ 24 or smaller is likely to originate from the smaller dynamic range associated with those data sets rendering the interpolation of the subset to fit the network input size more erroneous. To gain even deeper insights in the connection between dynamic range and the inversion results obtained from certain subsets, it is helpful to calculate the probability $P$ of some particular data points being included in the subsets considered. Straightforward calculations given in previous work of Beckmann \textit{et al.}\supercite{Beckmann_phd} result in the following expression:
\begin{equation} \label{eq:prob_us}
    \begin{split}
        P &= \sum_{i = 1} \sum_{j = 1} \binom{i + j}{i} \binom{n}{i + j} \\ 
        & \times \frac{N_{1}! \, N_{2}! \, N_{out}! \left(N - n\right)!}{\left(N_{1} - i\right)! \, \left(N_{2} - j\right)! \, \left(N_{out} - \left(n - \left(i + j\right)\right)\right)! \, N!},
    \end{split}
\end{equation}
where $N$ and $n$ are the total number of points of the fully sampled and sub-sampled signal respectively, $N_{1}$ and $N_{2}$ refer to the total number of points in some freely chosen subset of the original signal and $N_{out}$ is the total number of points not included in either of both subsets. If the sum of $N_{1}$ and $N_{2}$ fits the size of the sub-sampled signal, summation commonly runs up to $N_{1}$ and $N_{2}$. Otherwise, arbitrary summations limits can be imposed. Equation~\ref{eq:prob_us} calculates the probability, that a least one point of each of the subsets relating to $N_{1}$ and $N_{2}$ is included in the sub-sampled signal. If the chosen subsets further refer to $N_{1}$ smallest and $N_{2}$ largest values in the original signal, $P$ can be interpreted as a measure for the dynamic range of the sub-sampled signal with a high dynamic range coinciding with a probability close to unity. In the case of $N_{1}$ and $N_{2}$ being associated with the 16 smallest and largest values in the original 64 $\times$ 64 data set (coinciding with the 0.4\% smallest and largest values in the original signal), equation~\ref{eq:prob_us} yields for the highest level of random sub-sampling a probability of 0.05 and for the 32 $\times$ 32 subset P equals to 0.98, with an overall monotonically increasing trend converging towards 1 with increasing size of the subset.  The obtained probability trends show that the most strongly sub-sampled data sets (in particular the 8 $\times$ 8 and 16 $\times$ 16) cannot be expected to cover the full dynamic range of the original signal and consequently, the simultaneous drop in network performance further highlights the significance of dynamic range on the inversion results obtained via deep learning. Focusing on the combination of truncation and random sampling, comparable results to random sampling alone are achieved with the exception of a significant decrease in inversion performance in particular for MTGV and Tikhonov regularization, if the 8 $\times$ 8 or 16 $\times$ 16 subsets are employed. From initial intuition, it seems likely that truncation and random sampling is expected to result in a lower dynamic range of the generated subset compared to random sampling alone, because a considerable number of points in the subset is used for truncation and hence, not available for random allocation anymore. However, calculating the likelihood of the 8 $\times$ 8 subset to include at least one point from each of the 16 smallest and largest values of the original signal via equation~\ref{eq:prob_us} gives a probability of 0.17, which is more than three times as likely as random sampling alone. This is the case, because due to truncation the 16 largest data points are already included in every subset and therefore, increasing the chance that the remaining randomly sampled points generate a final subset which contains at least one point stemming from the 16 smallest values of the original signal. Hence, the combination of truncation and random sampling is likely to result in subsets with a generally higher dynamic range than it can be expected from random sampling alone. This means the observed drop in inversion performance is unlikely to originate from low dynamic range of the employed subset, instead it indicates that randomness of the created subsets is a further factor influencing reconstruction result. In more detail, the combination of truncation and random sampling generates subsets which are not fully random compared to data sets from random sampling alone. Under the consideration of the latter observation and the usually larger dynamic range of data sets generated via the combination of truncation and random sampling, it is eventually more likely that the drop in inversion performance originates from the reduced randomness of the employed subsets rather than being triggered by an insufficient dynamic range. Moving to the differences between the reconstructions shown in figure~\ref{fig:dis_rec}, the artefact peaks obtained for Tikhonov regularization are a well-established finding in the literature\supercite{Mitchell2012NumericalDimensions, Reci2017ObtainingProblems, Reci2017RetainingExperiments.} and stem from the smoothness enforcing mathematical properties of the $\mathrm{L_2}$-norm. Similarly, the overly narrow peaks in the reconstruction generated through the combination of Tikhonov and deep learning is an unsurprising finding considering the neural network employed. The neural network was trained as a filter using reconstruction results from Tikhonov regularization as an input and subsequently, generating an output distribution freed from regularization artefacts. This procedure enables the successful removal of reconstruction artefacts, but further imposes an inherent bias into the network. In more detail, Tikhonov regularization does not only generate artefact peaks, but occasionally tends to result in reconstructions with a considerably broader peak form than the original distribution. Using those reconstructed distributions as inputs for network training in combination with the original distributions as the training outputs results in a biased set of network weights, which efficiently removes additional artefact peaks but also imposes an additional narrowing of the major peaks even if Tikhonov regularization alone already obtained a close-to-original peak form. Moving to the $\phi$-score results, the reversed trend of the combination of Tikhonov and deep learning performing the best for fully sampled signals compared to receiving the worst $\chi$-score ranking shows that a fully objective and unbiased scoring metric is usually not achievable and that the best possible scoring method has to fit the purpose of application. Within this context, equation~\ref{eq:score_us_art} is designed to penalize any reconstruction artefacts very heavily and therefore, ensuring that every reconstruction including any kind of ambiguous peaks receives a poor ranking. Hence, the fact that Tikhonov regularization alone receives the worst $\phi$-score rankings, whereas its combination with deep learning ranks the best, can be considered as an expected outcome due to the high penalty for any form of artefact peaks enshrined in equation~\ref{eq:score_us_art}. This is also consistent with the methodological principles the combination of Tikhonov and deep learning is based upon. Methodologically speaking, this combined approach utilizes Tikhonov regularization results as inputs to a neural network designed to remove numerical artefacts, eventually providing an improved reconstruction compared to Tikhonov alone. The clear disparity between the $\phi$-score rankings received for the combination of Tikhonov and deep learning and its $\chi$-score analogue further highlights the importance of the ranking method being fit for purpose. Furthermore, the high sensitivity regarding artefact peaks is likely to be the reason for the generally more erratic behaviour of the $\phi$-score compared to the results received from equation~\ref{eq:score_us}. In more detail, different levels of sub-sampling result in a varying number and intensity of inversion artefacts which are very heavily penalized by equation~\ref{eq:score_us_art}. Consequently, the obtained $\phi$-scores show significant fluctuations even if the main features of the distribution are reconstructed with similar or close-to-equal accuracy. Comparing different sampling methods, similar overall trends compared to equation~\ref{eq:score_us} can be identified. Analogously to the $\chi$-rankings, the truncation results highlight the importance of high dynamic range for receiving a high quality reconstruction. Further in agreement with the observations made for equation~\ref{eq:score_us}, $\phi$-scores reveal a sharp drop in inversion performance at high sub-sampling levels if random sampling or its combination with truncation is employed. This trend is even more pronounced \hl{than is} the case for equation~\ref{eq:score_us}, indicating that severe sub-sampling and the associated reduced dynamic range of those subsets results in additional numerical artefacts which then are heavily penalized by equation~\ref{eq:score_us_art} explaining the observed drop in inversion performance. 

\section{Conclusion}
\label{sec:con_us}

Overall, the findings presented in this publication agree well with the trends observed for fully sampled signals from Beckmann \textit{et al.}. Comparing sub-sampling methods, it becomes evident that random sampling alone provides the best overall results independent of the inversion method, further emphasizing the major importance of dynamic range as well as randomness for the successful inversion of sub-sampled signals. Moving to a comparison between inversion techniques, deep learning shows the overall best performance if at least a quarter of the original signal is sampled and further provides the best compromise between the suppression of numerical artefacts and an accurate reconstruction of the original peak form. The reversed relative ranking positions obtained for the combination of Tikhonov and deep learning highlights the significance of choosing a ranking function which penalizes accordingly to the purpose of application. For instance, in the majority of experimental settings the ground truth distribution is unknown and hence, the suppression of inversion artefacts is of upmost importance for the unambiguous interpretation of experimental results. Consequently, in those cases equation~\ref{eq:score_us_art} is the more suitable scoring metric. However, if the ground truth distribution is known and an accurate estimate of the integral of a certain peak is the main experimental target, equation~\ref{eq:score_us} can be expected to be the more appropriate ranking method. In consequence, this also means that the choice of inversion algorithm eventually depends on the area of application and a final recommendation can only be made under consideration of the purpose of the experiment. From this it becomes evident that conducting similar simulations to the ones presented in this paper prior to an experiment establishes a sensible rational for choosing an optimal degree of sub-sampling enabling significant experimental time savings compared to the fully sampled signal. Finally focusing on Tikhonov regularization, the obtained results show that all things considered Tikhonov alone achieves the worst inversion performance of the methods investigated in this study. Consequently, the main argument in favour of Tikhonov regularization remains its speed advantage over MTGV and its robustness at high levels of sub-sampling. However, if neither extreme sub-sampling or a very fast inversion time is necessary, Tikhonov regularization has to be considered as the inferior inversion technique, questioning the fact that it is still the most widely used inversion method for a vast majority of experiments in the field of NMR.